\documentclass[]{spie}  %>>> use for US letter paper
\usepackage{amsmath,amsfonts,amssymb}
\usepackage{graphicx}
\graphicspath{{figures/}}
\usepackage[colorlinks=true, allcolors=blue]{hyperref}
\usepackage{caption}
\usepackage{subcaption}
\usepackage{comment}

\title{First Experimental Results of the Fast Atmospheric Self-coherent Camera Technique on the Santa cruz Extreme Adaptive optics Laboratory Testbed: Demonstration of High Speed Focal Plane Wavefront Control of Residual Atmospheric Speckles}

\author[a,b]{Benjamin L. Gerard}
\author[b,a]{Daren Dillon}
\author[c,b]{Sylvain Cetre}
\author[b,a]{Rebecca Jensen-Clem}
\author[b]{Thomas D Yuzvinsky}
\author[b]{Holger Schmidt}

\affil[a]{University of California Observatories, CA, USA}
\affil[b]{University of California Santa Cruz, CA, USA}
\affil[c]{W.M. Keck Observatory, HI, USA}

\authorinfo{Send correspondence to Benjamin L. Gerard: blgerard@ucsc.edu}

\begin{document} 
\maketitle

\begin{abstract}
Current and future high contrast imaging instruments aim to detect exoplanets at closer orbital separations, lower masses, and/or older ages than their predecessors, with the eventual goal of directly detecting terrestrial-mass habitable-zone exoplanets. However, continually evolving speckles in the coronagraphic science image still limit state-of-the-art ground-based exoplanet imaging instruments to contrasts at least two orders of magnitude worse than what is needed to achieve this goal. For ground-based adaptive optics (AO) instruments it remains challenging for most speckle suppression techniques to attenuate both the dynamic atmospheric and quasi-static instrumental speckles. We have proposed a focal plane wavefront sensing and control algorithm to address this challenge, called the Fast Atmospheric Self-coherent camera (SCC) Technique (FAST), which enables the SCC to operate down to millisecond timescales even when only a few photons are detected per speckle. Here we present preliminary experimental results of FAST on the Santa Cruz Extreme AO Laboratory (SEAL) testbed. In particular, we illustrate the benefit ``second stage'' AO-based focal plane wavefront control, demonstrating FAST closed-loop compensation of evolving residual atmospheric turbulence on millisecond-timescales.
\end{abstract}

% Include a list of keywords after the abstract 

\section{INTRODUCTION}
\label{sec:intro}  % 
Current state-of-the-art instrumentation enables ground-based exoplanet imaging instruments to be sensitive to detecting self-luminous giant exoplanets down to a few Jupiter masses at separations beyond around 10 au within star systems younger than a few hundred Myrs\cite{gpies}. Despite the exciting new detections enabled from these instruments\cite{51eri}, large surveys have both (1) shown that the giant exoplanets these facilities are sensitive to detecting are rare---as low as 1\% of stars hosting such planets\cite{idps}---and as a result (2) illustrated that future instruments will need to improve detection and characterization sensitivity to lower mass, closer-in, and older systems. The main technological factor limiting such improved sensitivity is un-corrected/subtracted speckle noise. Speckle noise is a result of leftover diffracted starlight in the coronagraphic science image that prevents detecting exoplanets below an astrophysical flux ratio (i.e., planet flux normalized to the host star flux), typically around $10^{-6}$ at radial separations around 10$\lambda/D$ from the star\cite{fast_phd}. It is therefore crucial to improve on these speckle subtraction and correction techniques to enable both current and the next generation of exoplanet imagers to detect and characterize new exoplanetary systems, shedding light on the processes of planet formation, evolution, and ultimately the prevalence of life beyond the Solar System\cite{psi}.

Focal plane wavefront sensing and control\cite{cdi_rev}---using either an active feedback loop between the coronagraphic science image and adaptive optics (AO) system's deformable mirror (DM) and/or via post-processing methods, called coherent differential imaging (CDI)---is one such speckle correction technology that could enable the aforementioned sensitivity/contrast gains needed for fainter exoplanet detection and characterization. However, minimal on-sky gains with this approach have been demonstrated thus far\cite{scexao_nulling, scexao_ldfc}, largely due to speckle evolution occurring on timescales faster than measurement and correction algorithms and/or system hardware can enable\cite{fast_phd}. However, a plethora of projects are now being pursued to further develop the promise of this technology on-sky, including ongoing efforts at Keck/NIRC2\cite{keck_speckle_nulling}, Subaru/SCExAO\cite{scexao_ldfc,scexao_mec}, Magellan/MagAO-X\cite{magao_cdi}, ESO/SPHERE\cite{sphere_efc}, and Gemini/GPI\cite{cal2}. In this paper we will discuss related laboratory developments of one such technology, based on the self-coherent camera (SCC)\cite{scc_orig}, called the Fast Atmospheric SCC Technique (FAST) \cite{fast18,fast_phd}, using the Santa cruz Extreme AO Laboratory (SEAL)\cite{seal}. Note that this FAST concept was initially developed at the high contrast testbed at the National Research Council of Canada Herzberg Astronomy and Astrophysics research center, which has published laboratory optical design and preliminary in-house FAST testing results in Ref. \citenum{ne}. We summarize FAST in \S\ref{sec: fast_sum}, provide an overview of our FAST SEAL setup in \S\ref{sec: setup}, and describe various FAST hardware characterization and calibration procedures in \S\ref{sec: fpm}-\ref{sec: calib}. We present results and analysis of FAST correction and CDI of quasi-static aberrations in \S\ref{sec: quasi_stat}. In \S\ref{sec: dynamic} we present the main results and analysis of this paper: real-time FAST correction of AO residual turbulence, clearly demonstrating the gain that high-speed focal plane wavefront control enables. We then provide further discussion in \S\ref{sec: discussion} and conclude in \S\ref{sec: conclusion}.
\section{FAST SUMMARY}
\label{sec: fast_sum}
\begin{figure}[!h]
\centering
\includegraphics[width=0.98\textwidth]{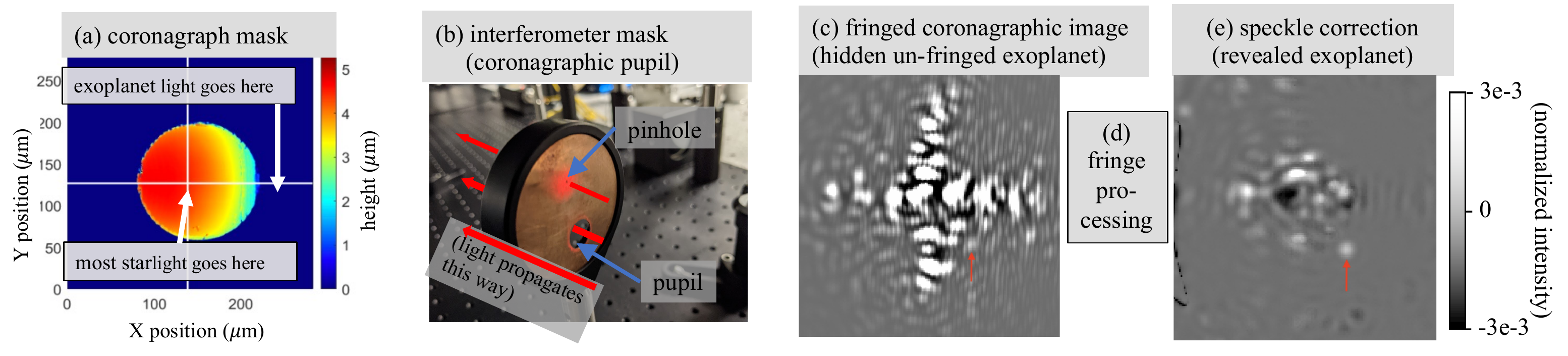}
%\vspace*{-10pt}
\caption{Overview of FAST, using images from our recently-assembled Santa cruz Extreme AO Laboratory (SEAL) high contrast testbed\cite{seal} (these FAST/SEAL laboratory results will be presented throughout this paper and are shown here to first illustrate the FAST concept). A custom coronagraph mask (a) is designed to both (i) attenuate starlight while transmitting exoplanet light, and (ii) enhance the stellar fringe visibility using an interferometer mask in the downstream coronagraphic pupil (b). This design enables the detection of an interference pattern on top of the leftover stellar speckles in the coronagraphic image, even when only a few photons are detected per pixel (c), enabling coronagraphic science imaging and AO wavefront sensing. The fringed stellar speckles are then processed (d) and further attenuated using both a deformable mirror and via post-processing  (e), revealing an un-fringed exoplanet. See \S\ref{sec: closed_loop} for more details on panels d and e.\vspace{10pt}}
\label{fig: fast_sum}
\end{figure}

FAST, illustrated and described in Fig. \ref{fig: fast_sum}, relies on a technique in which coherent starlight is interfered with itself to enable a measurement and correction of the complex speckle electric field (i.e., phase and amplitude) in the coronagraphic image. Speckle attenuation/subtraction with this approach can be accomplished using a DM (i.e., focal plane wavefront control)\cite{fast_spie18}, and/or by post-processing (i.e., CDI)\cite{fast18}.
As shown in Fig. \ref{fig: fast_sum}, FAST produces stellar interference fringes by way of two specialized coronagraphic masks: a custom focal plane mask (Figure \ref{fig: fast_sum}a) modulates starlight but not exoplanet light such that the downstream pupil plane mask (Figure \ref{fig: fast_sum}b) transmits remaining starlight and exoplanet light through a traditional Lyot stop, while much of the starlight (but not exoplanet light) passes through a separate pinhole in the mask. In the final focal plane (the coronagraphic image), the residual stellar speckles interfere with the starlight that passed through the pinhole to produce fringes (Figure \ref{fig: fast_sum}c) while any off-axis sources remain un-fringed). Fourier-based processing of the fringes in this final image enable estimating and attenuating (via DM control and/or CDI-based post-processing) remaining stellar speckles from a single, millisecond-exposure image, with no need to create additional phase diversity by defocusing and/or probing individual speckles with the DM (Figure \ref{fig: fast_sum}e). 

Reviewing a mathematical description of the SCC/FAST image, the recorded image on the focal plane detector, $I$, produced from noiseless (i.e., without photon/detector/sky background noise) propagation of wavefront error (WFE; i.e., from atmospheric and/or instrumental origin) through to the detector, can be described by\cite{scc_orig}
\begin{equation}
    I=I_S+I_P+I_R+2\sqrt{I_S I_R}M,
\end{equation}
where $I_S$ is the un-fringed stellar speckle intensity component, $I_P$ is the un-fringed planet intensity component, $I_R$ is the pinhole point spread function (PSF) component (i.e., the PSF recorded if the main pupil aperture was blocked), and $M$ is a dimensionless fringe variable that varies between 0 and 1. The fringe term $2\sqrt{I_S I_R}M$---which can be isolated in the Fourier plane of the image (known as the optical transfer function or OTF, the amplitude of which is known as the modulation transfer function or MTF)---encodes the coronagraphic image intensity ($\sqrt{I_S I_R}$) and phase ($M$) in a single image without bias from the un-fringed exoplanet signal $I_P$. The phase of a given stellar speckle, which is not normally measureable in a single cornagraphic image, is encoded by the relative position of fringes (i.e., within $M$) on that speckle, which changes with the differential electric field phase at the detector focal plane between the $I_S$ and $I_R$ components (for which those terms alone only represent the square modulus of the electric field and do not include phase information).

FAST's key innovation compared with previous focal plane wavefront control and CDI techniques is the coronagraphic focal plane mask (Fig. \ref{fig: fast_sum}a), which provides a high enough fringe visibility (i.e., fringed vs. un-fringed components) in the final image such that fringes can be detected on millisecond timescales, even when only a few photons are detected per pixel. FAST's predecessor, the ``self-coherent camera," also generated fringes in the coronagraphic image to enable focal plane wavefront control and CDI \cite{scc_orig}, but with fringe visibilities about 10$^6$ times lower than Fig. \ref{fig: fast_sum}c \cite{fast_phd}, removing any possibility of AO-based wavefront control on millisecond timescales.
\section{LABORATORY SETUP}
\label{sec: setup}
In this section we will describe our refractive setup developed to test FAST, which is the same setup from which subsequent results in this paper are obtained. Although we will refer to this setup as the Santa cruz Extreme AO Laboratory (SEAL), SEAL will ultimately be a mostly reflective setup designed for multi-purpose AO-based wavefront sensing and control techniques, beyond the scope of just testing FAST; the higher-level setup of the full SEAL testbed, including the optical design and both current and ongoing projects (including FAST),  are presented and described in Ref. \citenum{seal} from this same conference proceedings. In this paper we will instead focus only on the results obtained from the dedicated FAST setup, although the goals here are analogous to those for testing FAST on the full SEAL testbed. Note that the main facilities used here are the same as previously described in Ref. \citenum{lao_granite}, including a highly-stabilized granite testbed and custom-made testbed enclosure.

\begin{figure}[h]
\centering
\includegraphics[width=1.0\textwidth]{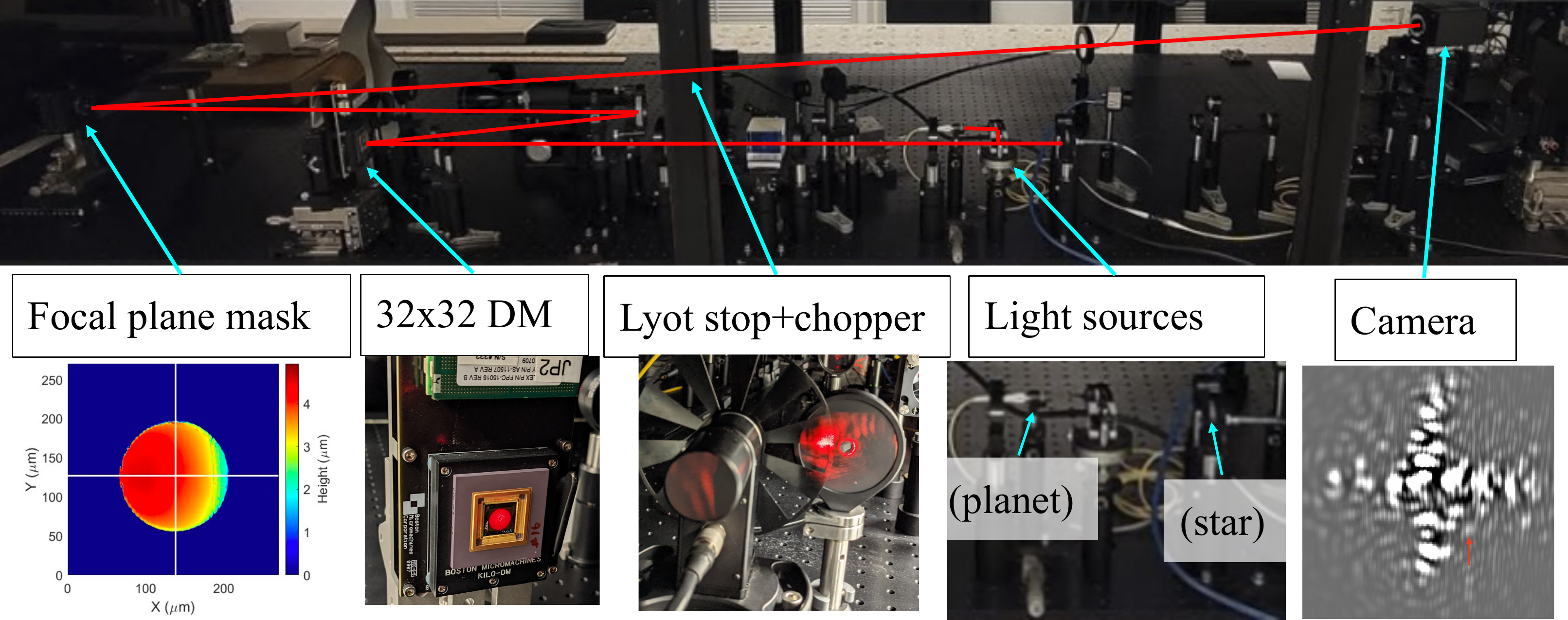}
\caption{An outline of our FAST SEAL setup. A planet and star light source, both aligned to the same focal plane, are combined and then collimated, with the planet source positioned slightly off-axis a few diffraction bandwidths from the star. An aperture stop and re-imaging optics place the pupil on a DM, sampling 29 actuators across the beam. The beam is then focused on our custom FAST focal plane mask, then re-collimated to a downstream coronagraphic pupil, where a Lyot stop and optical chopper modulate the beam. Finally, downstream imaging optics enable coronagraphic focal plane (i.e., FAST) or pupil plane imaging (the chopper and Lyot stop are on repeatable magnetic mounts to enable un-occulted or occulted coronagraphic pupil imaging).}
\label{fig: lab_setup}
\end{figure}

Fig. \ref{fig: lab_setup} shows our FAST SEAL setup and outlines the main hardware components.
The visible light star and planet light sources are commercial Thorlabs lasers KLS635 and HNLS008L, respectively, centered at 633 nm. The adjustable star light source is set to 0.15 mW unless otherwise noted, while the planet light source is fixed at 0.8 mW. We place a neutral density (ND) filter with an optical density (OD) of 3 on the planet source\footnote{Note that we found that adding the ND filter to the planet light source measurably changed its optimal focus position, requiring manual alignment with the ND filter on the planet source and the star light source turned off.}. The beamcube combining the two light sources further attenuates the planet by a factor of 9, theoretically setting the planet at an astrophysical flux ratio of 6$\times10^{-4}$ (see \S\ref{sec: calib} for a measurement of the star-to-planet flux ratio). Additional higher OD ND filters (and/or increasing the star light source power) can further decrease this flux ratio for future high contrast experiments, but for the purposes of our main focus in this paper (i.e., high speed FAST correction of AO residuals) the above-described planet flux ratio is sufficient. Moving downstream, a pupil stop defines the system aperture. The beam cube in Fig. \ref{fig: lab_setup} just downstream is used for a separate experiment with our spatial light modulator (SLM), described in Ref. \citenum{seal}. A set of relay lenses then re-images the pupil onto our 32 x 32 actuator microelectricalmechanical system deformable mirror (MEMS for short), illuminating 29 actuators across the beam diameter (see Ref. \citenum{lao_mems} for a detailed description and characterization of this device). Re-imaging optics then relay to an empty pupil plane populated by a fold mirror, initially planned to be replaced by our 97 actuator ALPAO DM\cite{seal}, but ultimately not needed for these FAST tests. A plate beam splitter then generates a separate path which after relay optics images the pupil on a Thorlabs Shack Hartmann wavefront sensor (SHWFS; model WFS-20). Although in this paper we do not use the SHWFS in tandem with FAST real-time operations, this setup has helped to enable software development to subsequently enable this in future tests; below we briefly discuss such testing with this SHWFS. A 500mm focal length lens then generates a f/37 beam on the focal plane mask (FPM), which is designed and fabricated specifically for that f-ratio and the 633 nm light source wavelength. A f=200mm, 2"{\O} lens then collimates the post-FPM beam, oversizing the clear aperture relative to the non-coronagraphic pupil footprint by a factor of 6.2, sufficient for the ``classical'' SCC\cite{galicher_scc}. A Lyot stop then transmits light through the central 7.3mm coronagraphic pupil (90\% undersized) and off-axis 0.45 mm pinhole (which is the maximum pinhole diameter required to enable the first pinhole PSF Airy minimum to lie outside the DM control region; see equation 3.2 in Ref. \citenum{fast_phd}). Although the FAST FPM prescription is designed for a theoretical pinhole-pupil separation (center to center) of 1.6 pupil diameters, the separation is instead empirically determined to be 1.524 pupil diameters (see \S\ref{sec: fpm_alignment}). Our optical chopper and controller are the off-the-shelf MC2000B model from Thorlabs, using the MC1F10 blade. A f=500mm 2"{\O} lens lastly focuses the post-Lyot stop beam onto our Andor Zyla 5.5 sCMOS camera, enabling a theoretical plate scale of 6 pixels/resel (see \S\ref{sec: calib_ini} for a corresponding measurement), sufficiently oversampled for the SCC fringes\cite{galicher_scc}. A f=50mm lens is placed 50mm from the detector on a flip mount to reimage the coronagraphic pupil on the same detector when deployed. The Lyot stop is also on a flip mount to enable coronagraphic pupil imaging both with and without the Lyot stop.

Our software infrastructure is based on the Keck pyramid wavefront sensor real time control (RTC) architecture, described in Ref. \citenum{krtc}. Ref. \citenum{seal} describes more specifically the hardware and infrastructure for SEAL software control and development, which includes a GPU-based clone of the Keck RTC in Linux (which we will refer to as SEAL) and a Windows machine for some off-the-shelf components without available Linux drivers. Low-level drivers for the MEMS and Andor camera (i.e., to get and send DM commands and to get images/sub-arrays) are written on SEAL in C and interfaced with Python to enable high-level FAST software development. Unless otherwise mentioned we use a 320x320 sub-array updating at 100 Hz (although in future work we will plan to demonstrate the higher-speed FAST closed-loop control: Andor benchmarks the Zyla 5.5 at less than 1 e- read noise at up to 1 kHz frame rates; Andor, private communication). Running in serial mode in Python (i.e., not parallelized), a 20 ms pause between grabbing Andor frames and applying DM commands is needed to enable a given image to be used to compute the corresponding DM commands; although we are working to decrease this latency by multi-threading and synchronizing imaging-grabbing and DM-command-sending threads, we will ultimately implement a high-speed high-level FAST RTC in C, analogous to the above-described Keck pyramid wavefront sensor RTC and enabling computational latencies closer to 1 ms (consistent with FAST CPU-based latencies computed in Ref. \citenum{fast_spie18}). Our windows machine is used to operate the adjustable star light source, deployable pupil imaging lens, optical chopper wheel, and SHWFS. The optical chopper wheel is electronically synchronized to the Andor camera readout frame rate, using the Andor as the ``leader'' and the chopper as the ``follower,'' manually adjusting the chopper phase to ensure the Lyot stop pinhole in every other frame is blocked and then unblocked. For the SHWFS, although Thorlabs does not provide Linux-based drivers for the commercial WFS-20, we use the Windows serial interface in Python and an ethernet connection to SEAL to provide high speed SHWFS frames in our SEAL Python interface for high-level AO software development (as with the Andor and MEMS above, real-time code will ultimately be converted into C to enable optimal high speed performance). Benchmarking with this SHWFS setup shows that we can readout slopes with 26 subapertures across the beam (i.e., only slightly undersampled relative to the 29 DM actuators across the beam) at 460 Hz for future ``first stage AO + FAST'' tests at high speed (see \S\ref{sec: conclusion}).
\section{CORONAGRAPH MASK CHARACTERIZATION}
\label{sec: fpm}
\subsection{FPM and Lyot stop alignment}
\label{sec: fpm_alignment}
Off-the-shelf lenses are aligned using a shear plate to collimate the beam and a knife-edge test to calibrate focal lengths but will not be discussed in detail further in this paper. Here we will further discuss the alignment procedure for the custom FAST FPM and Lyot stop.

FAST FPM alignment is carried out in multiple steps: initial coarse adjustment with a 5 degree-of-freedom stage (x-y-pitch-yaw-roll; focus is pre-determined by a knife edge test as described above). With the Lyot stop in place, pitch and yaw (i.e., tip and tilt) of the FPM mount are adjusted until the non-coronagraphic pupil aperture is centered on the intended central aperture of the Lyot stop (which displays a 0.8mm-wide ring around the edge due to the 90\% undersized aperture). Then a pinhole aperture is placed just upstream of the FPM to simulate a flat field. The FPM core in this mode is visible as a dark circular shadow on the otherwise illuminated flat field; x and y positions are then adjusted so that the center of this black dot is positioned at the coordinates of the normal Airy disk's center. After removing the pinhole/flatfield, a somewhat-aligned coronagraphic pupil is visible, by eye, on the Lyot stop (as in Figures \ref{fig: fast_sum} and \ref{fig: lab_setup}, but not as well- aligned yet); further x-y adjustment can then enable visually maximizing the off-axis pupil intensity into the pinhole. The FPM roll is then adjusted to the desired position angle (i.e., defined by the Lyot stop pinhole's position angle), after which the above x-y positioning steps are repeated (since the custom shape on the FPM is not exactly in the center of the optic's clear aperture). Lastly, fine x-y adjustment is performed in software with the DM tip/tilt: a grid search is performed in a square region around the estimated best alignment position, optimizing a combination of maximal fringe visibility \textit{and} minimal raw contrast.\footnote{Note: although intuitively it might seem like fringe visibility alone would be a sufficient metric to align the FAST FPM, since once fully aligned the maximal amount of light is transmitted through the Lyot stop off-axis pinhole, fringe intensities are equally weighted between the focal plane amplitudes from the Lyot stop pinhole \textit{and} pupil (see eq. 3.1 in Ref. \citenum{fast_phd}), which causes fringe intensities to increase when the FPM is misaligned because the amount of starlight increase is greater than the amount of pinhole light decrease. This is mitigated as described above by requiring minimal starlight in addition to maximal fringe visibilities (as measured by integrated flux on the modulation transfer function sidelobe).} 

The pupil-to-pinhole separation for our Lyot stop prescription is determined empirically, as was done by Ref. \citenum{fast_ne}, by taking coronagraphic pupil images with the Lyot stop out and fitting the centers of each pupil (where the DM is replaced by a fold mirror for this characterization to enable a more-diffraction-limited system; also see the next two subsections). The the off-axis pupil center is determined by fitting a two-dimensional Gaussian, while the central coronagraphic pupil position is determined by eye (incurring $\sim$1-2 pixel error $\approx$  2\% pupil diameter error $\approx$ 16 \% pinhole diameter error). Measuring this separation is crucial to enable a Lyot stop prescription that both (1) optimizes fringe visibility (i.e., limiting the pinhole position to deviate from the peak intensity), and (2) prevents a differential tilt between the pinhole PSF center and unfringed coronagraphic image center (i.e., preventing an offset between the center of the fringed and unfringed image components). Although we discuss the latter topic further in \S\ref{sec: tilted_pinhole_PSF}, we initially found that a Lyot stop prescription with a pinhole separation too discrepant from what was needed indeed caused a tilted pinhole PSF with respect to the unfringed coronagraphic image due to an amplitude gradient across the pinhole aperture causing a tilt. Although our initial pinhole separation prescription was for the intended FPM design, as discussed in \S\ref{sec: setup} it turned out that this differed significantly enough from the fabricated mask to cause such a measureable descrepancy, but with an empirically measured Lyot stop prescription this was no longer a problem. A recorded coronagraphic pupil image should therefore be the main determinant for future SCC Lyot stop prescriptions, both for lab experiments and future FAST instruments.
\subsection{Tip-tilt Gaussian mask}
\label{sec: uoa_fpm}
Fig. \ref{fig: fast_sum}a and \ref{fig: lab_setup} already show the FAST FPM fabricated and characterized via Zygo interferometry measurements by University of Alberta's nanoFAB laboratory, made from a Nanoscribe 3D printing machine and subsequently aluminum coated. This mask design only deviates from a flat surface over the central region that redistributes the core starlight, and with this 3D printing approach the full tilt can be made without phase wrapping (i.e., avoiding additional related chromatic effects). However, the design is also essentially a Lyot coronagraph, with no component optimized for diffracted starlight suppression, and so it is not ideal for reaching high contrasts. However, such a mask is still sufficient for the focus of this paper (i.e., ground-based observations where coronagraphic images are dominated by un-pinned speckle noise).

\begin{figure}[!h]
    \centering
    \begin{subfigure}[b]{0.36\textwidth}
        \includegraphics[width=\textwidth]{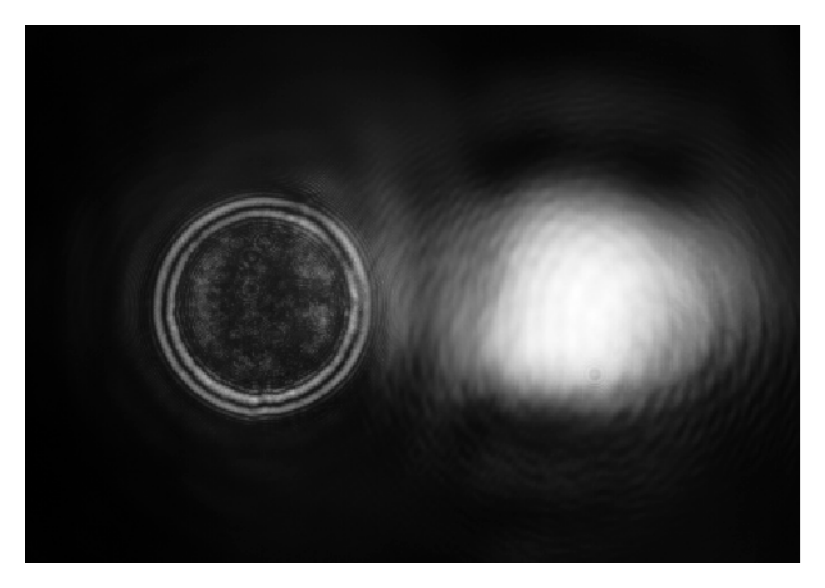}
        \caption{}
    \end{subfigure}
    \hfill
    \begin{subfigure}[b]{0.24\textwidth}
        \includegraphics[width=\textwidth]{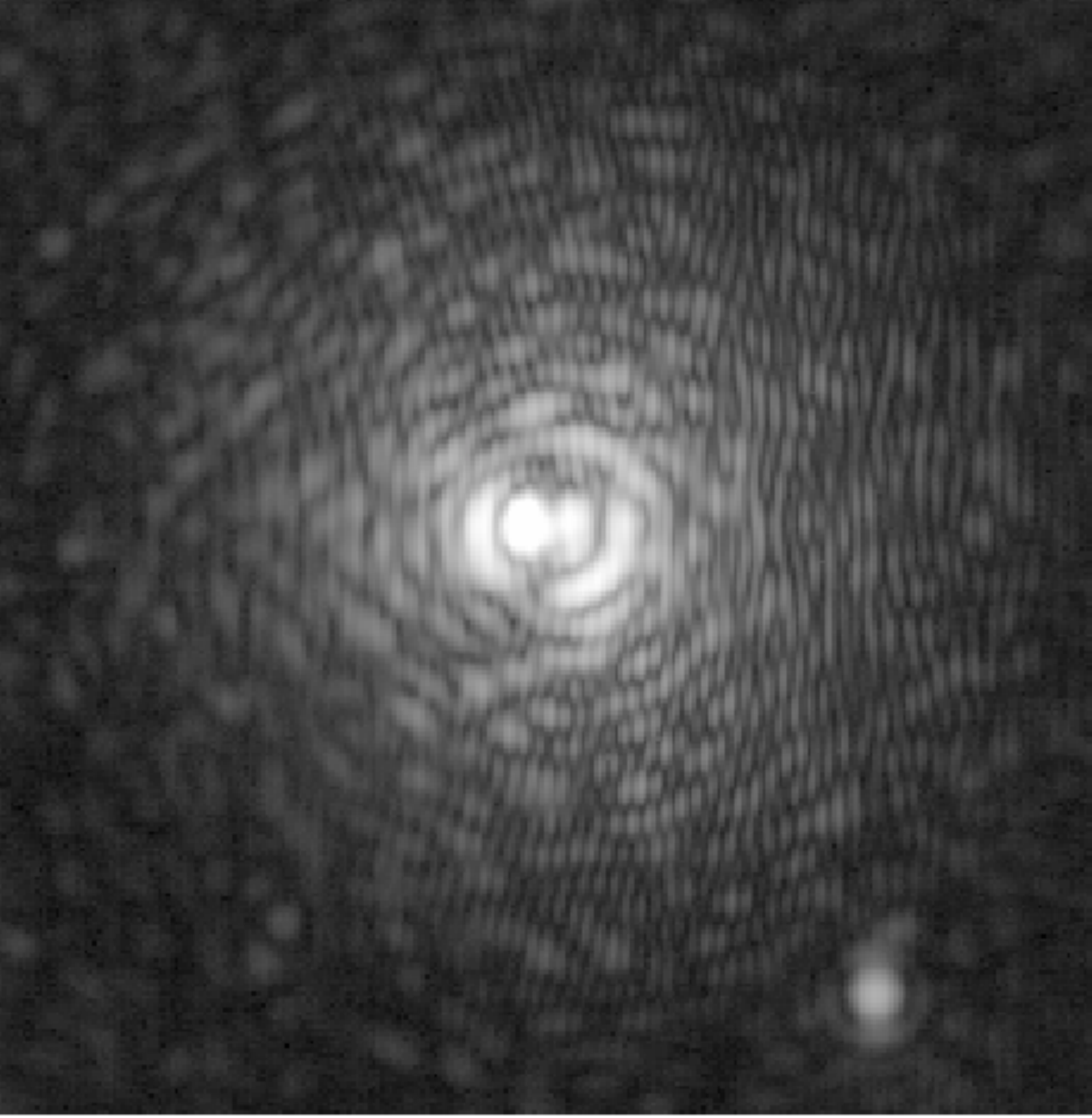}
        \caption{}
    \end{subfigure}
    \hfill
    \begin{subfigure}[b]{0.33\textwidth}
        \includegraphics[width=\textwidth]{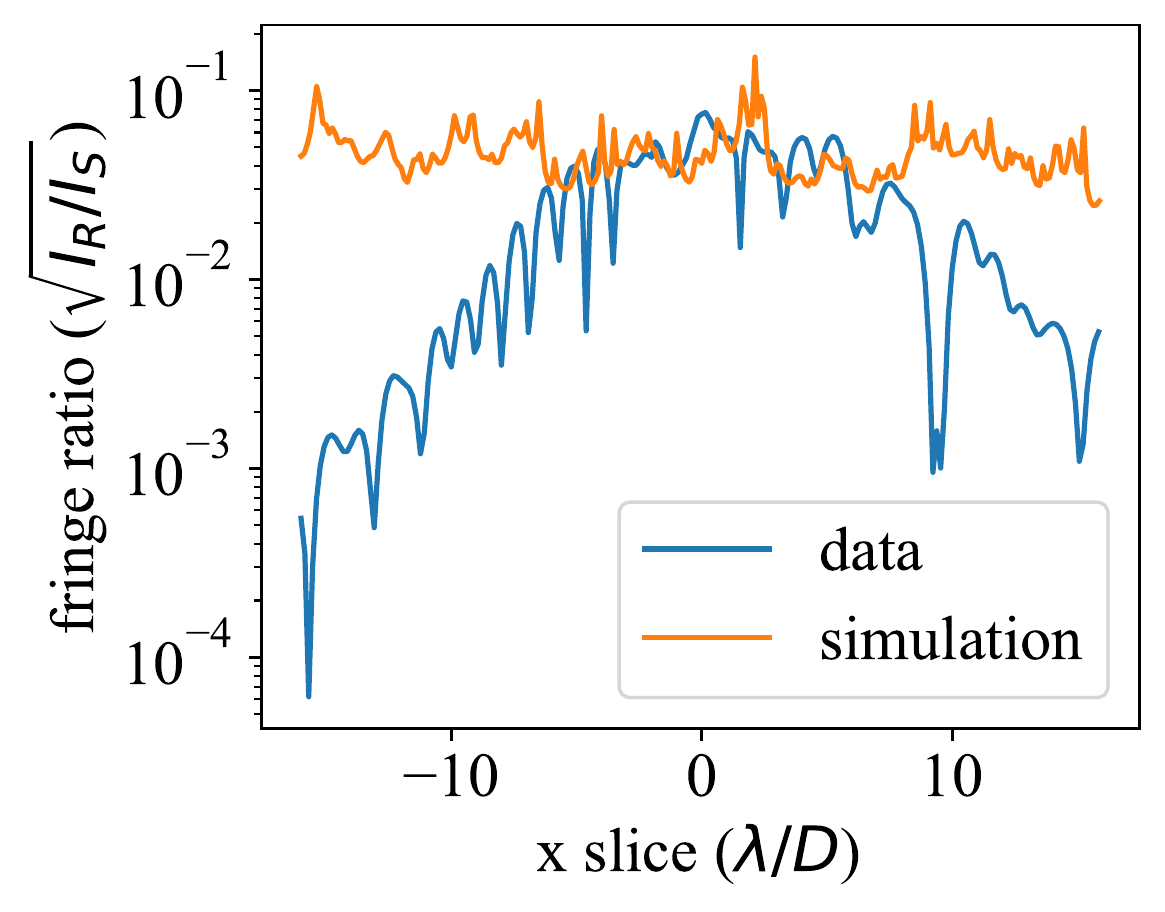}
        \caption{}
    \end{subfigure}
    \caption{Images and characterization of our FAST Tip/tilt Gaussian (TG) FPM made by University of Alberta's nanoFAB laboratory. (a) Coronagraphic pupil image (with the Lyot stop removed). (b) Coronagraphic focal plane image, with the Lyot stop in. Both (a) and (b) are shown on a logarithmic scale. (c) Fringe ratio analysis, showing the fringed ($I_R$) vs. unfringed ($I_S$) amplitudes for both the measured bench data and a simulation including both pupil and focal plane wavefront errors.  See the text for further simulation details.}
    \label{fig: uoa_fpm}
\end{figure}
Fig. \ref{fig: uoa_fpm} shows our initial characterization of mask performance, where in these tests the MEMS is replaced by a fold mirror. As shown, there is a clear off-axis pupil enabled by this FAST FPM in panel a, with fringes in panel b clearly detected and a pinhole PSF center co-aligned with the unfringed coronagraphic image center (see \S\ref{sec: tilted_pinhole_PSF}). Note that an optical ghost is present in the lower right of panel b, but because it is incoherent (i.e., unfringed) with starlight from the main beam, the SCC fringe processing algorithm is un-baised by such effects (i.e., there is no/negligible impact on measurement errors, although incoherent ghosts will still limit final achievable un-fringed contrasts). In panel c, the fringe ratio is a measurement of how bright the fringes are relative to the unfringed coronagraphic image component. This approach is similar to the first FAST FPM characterization carried out in Ref. \citenum{fast_ao4elt}. This is measured empirically from the lab data by Fourier filtering a MTF sidelobe of the SCC image, whose absolute value (i.e., ignoring phase) produces $\sqrt{I_R I_S}$. By separately recording an unfringed coronagraphic image ($I_S$, which we do using the optical chopper as described in \S\ref{sec: setup}), the fringe ratio image is computed by dividing the Fourier-filtered fringed image by the unfringed image:$\sqrt{I_R I_S}/I_S=\sqrt{I_R/I_S}$, for which Fig. \ref{fig: uoa_fpm}c shows a slice along the center of the image. The same process is repeated in simulation for our TG FPM prescription, where normalized WFEs projected in the entrance pupil and on the TG FPM are both 300 nm rms\footnote{assuming a -2 power law, normalized between 0 and 16 c/p} (larger than this would not be a diffraction-limited system), both averaged over 20 random wavefront realizations with the same wavefront rms to avoid bias from a single realization. Although this 300 nm rms level is within the expectation of overall system  and FPM fabrication capabilities, there is still a clear discrepancy at larger separations between simulations and lab measurements. Although this discrepancy should be investigated further, it is beyond the scope of this paper, which aims to demonstrate high speed wavefront control rather than characterize FPM fabrication to high simulation fidelity.
\subsection{Wrapped 4-level Tip-tilt Gaussian Vortex mask}
\label{sec: ucsc_fpm}
\begin{figure}[!h]
    \centering
    \begin{subfigure}[b]{0.31\textwidth}
        \includegraphics[width=\textwidth]{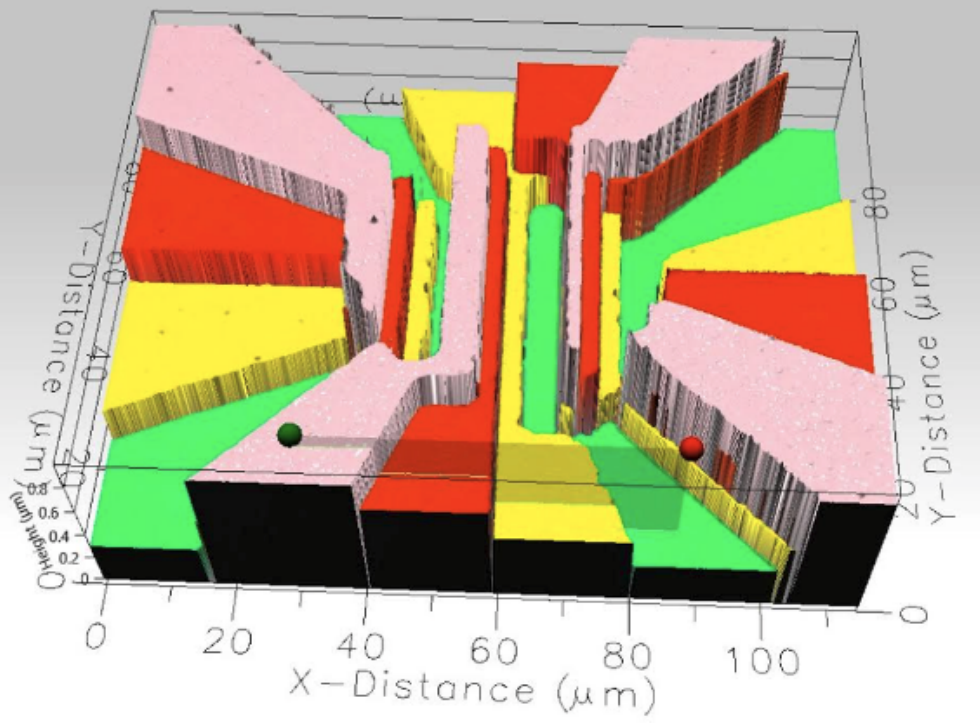}
        \caption{}
    \end{subfigure}
    \hfill
    \begin{subfigure}[b]{0.32\textwidth}
        \includegraphics[width=\textwidth]{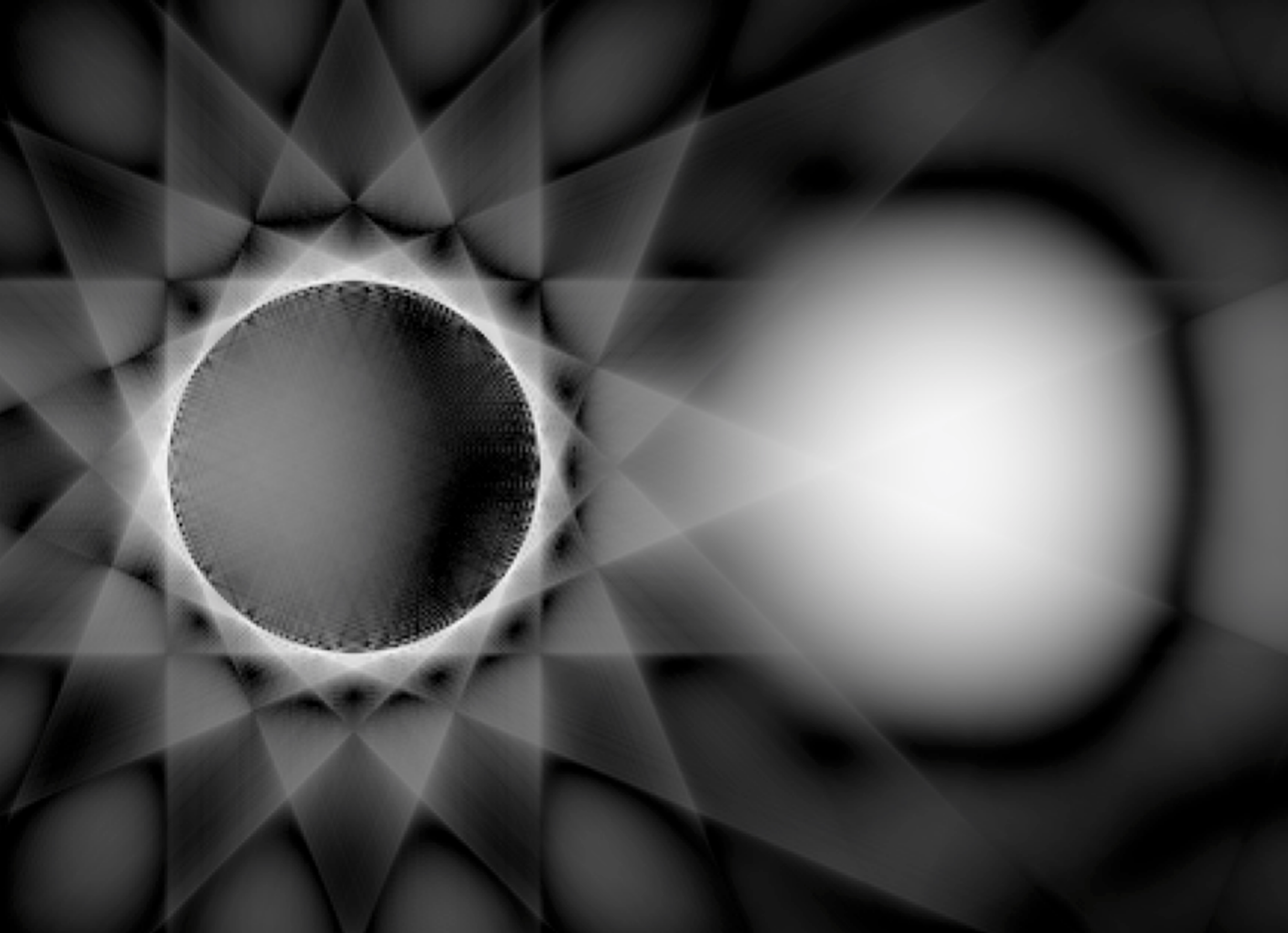}
        \caption{}
    \end{subfigure}
    \hfill
    \begin{subfigure}[b]{0.32\textwidth}
        \includegraphics[width=\textwidth]{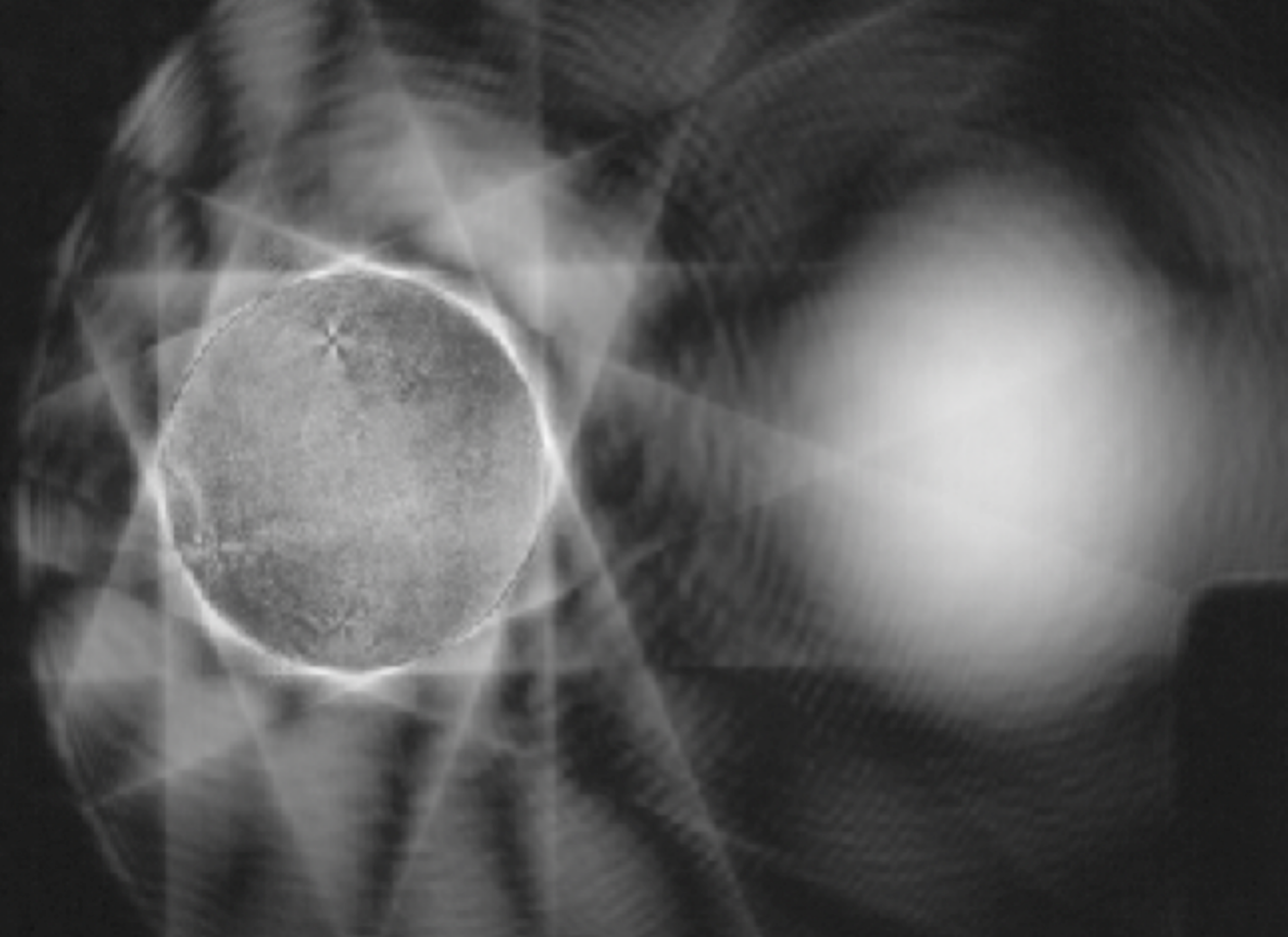}
        \caption{}
    \end{subfigure}
    \caption{Images and characterization of our FAST Tip/tilt Gaussian Vortex (TGV) FPM, fabricated by UCSC's Electrical and Computer Engineering nanofabrication facilities. (a) Topography of our four-level mask, obtained using an optical profilometer. (b and c) simulated and measured (respectively) coronagraphic pupil image (with the Lyot stop removed), both shown on a logarithmic scale. Note that as in \S\ref{sec: uoa_fpm}, in panel c the MEMS is replaced by a fold mirror to minimize overall system wavefront error for this characterization.}
    \label{fig: ucsc_fpm}
\end{figure}

We are also in the initial stages of development for fabricating FAST coronagraph masks at UCSC's W.M. Keck Center for Nanoscale Optofluidics. In contrast to outsourcing FAST mask fabrication through a vendor, development to support this collaboration internal to UCSC is advantageous to explore the capabilities and tolerances that can be achieved with both multiple mask designs and multiple fabrication techniques. We have thus far fabricated and tested one FAST mask, illustrated in Fig. \ref{fig: ucsc_fpm}, although more design and fabrication efforts are ongoing for future FAST developments and testing (see \S\ref{sec: future_ucsc_fpms}).

Figure \ref{fig: ucsc_fpm}a shows a surface depth profile characterization (via interferometric methods) of our FAST Tip/tilt Gaussian Vortex (TGV) coronagraph\cite{fast_tgv}, fabricated by four-layer titanium deposition. Unlike 3D printing the TG mask in \S\ref{sec: uoa_fpm}, here we must phase wrap the central TGV tilt angle due to the limited achievable dynamic range and number of layers with this ``lift-off" fabrication technique, adding additional chromaticity for a broadband FAST system but negligible for our monochromatic tests here\cite{fast_spie18}. Although a forthcoming manuscript will describe in more detail the fabrication process, in short it consists of (1) aligning a binary photomask with a photoresist-coated silicon wafer and exposing it to UV light, (2) puddle developing to remove exposed photoresist, (3) evaporating metal (in this case, Ti) across the entire surface,  followed by (4) lifting off the remaining photoresist, leaving behind patterned metal features of a single, uniform thickness, and finally (5) repeating steps 1-4 three more times to generate a four-layer scalar coronagraph mask. Comparing simulation to measurement in Fig. \ref{fig: ucsc_fpm}b and c, respectively, it is clear that we have fabricated a high-quality FPM with this approach; although it is difficult to give a quantitative overview without a simulated analysis in the coronagraphic image plane as in \S\ref{sec: uoa_fpm},\footnote{Due to timing and other laboratory testing constraints, we were not able to fabricate a Lyot stop mask based on the measured coronagraphic image in Fig. \ref{fig: ucsc_fpm}c and then carry out a fringe ratio analysis as in \S\ref{sec: uoa_fpm}.} various diffractive features in the coronagraphic pupil become less obvious at beyond a few hundred nm rms of FPM figure error\cite{fast_phd}, suggesting the quality is at least at this sufficient level. 
\section{CALIBRATION PROCEDURE}
\label{sec: calib}
\subsection{Initial Steps}
\label{sec: calib_ini}
%
%include generating a best flat with the Zygo and focal plane sharpening, planet contrast calibration measurement, finding the image center
Aligning the MEMS into the beam to use with FAST first required optical flattening to enable a diffraction-limited system; exposure to non-negligible humidity levels have slowly degraded this DM (purchased in the early 2000s\cite{lao_mems}) over time (also see \S\ref{sec: stability}), rendering the previously-generated best flat\cite{lao_mems} now aberrated by a few hundered nm of wavefront peak-to-valley. Although, initially we tried to flatten the DM using the SHWFS described in \S\ref{sec: setup} (which samples 29 lenslets across the 29 actuator beam), we ultimately opted to use our Zygo Interferometer\cite{lao_mems}, sampling over 1k pixels across the reconstructed wavefront. We took a synthetic interaction matrix approach to simplify processing across two computers (with the Zygo software on a separate Windows machine than the main SEAL Linux machine) by (1) recording the Zygo reconstructed wavefront for a push and pull of four separate actuators, (2) averaging the measured wavefront influence functions of each poke into a single influence function for a given DM actuator voltage applied, and then (3) assuming a system of uncoupled linear actuators in generating a synthetic interaction and command matrix. With this approach, we were able to converge at a 17 nm rms best flat command after 11 iterations. This also provided a calibration between DM units and reconstructed (reflected) wavefront, which we use subsequently in this paper: 1968nm of wavefront for 1 DM unit (i.e., the full stroke of the MEMS), consistent with the expected 1 $\mu$m (physical) stroke limit\cite{lao_mems}.

To calibrate contrast, with the best flat on the MEMS applied as described above we tilt the PSF off the FAST FPM core to record an off-axis PSF, adjusting the exposure time to prevent the star from saturating. Coronagraphic images in units of normalized intensity and/or contrast curves shown later in this manuscript are derived from dividing the dark-subtracted coronagraphic image in ADUs by the aforementioned maximum flux by the maximum of the (also dark-subtracted) off-axis stellar PSF, scaling to account for the different exposure times in the two images. We can also calibrate astrophysical flux ratio of the off-axis planet with this approach: turning off the starlight source but leaving on the planet source and scaling for exposure time gives a normalized planet flux ratio of 8.3$\times10^{-4}$.

To find the image center and empirically calibrate the plate scale we place two sine waves simultaneously on the DM--one with a position angle offset 90 degrees from the other and each at spatial frequencies of 10 c/p\footnote{Note that our SHWFS images confirm that 29 actuators are illuminated across the beam, which is assumed in order to convert a sine wave command applied on the DM into a wavefront spatial frequency.}--to generate ``satellite spots.'' Averaging the four 2D Gaussian-fit positions of these spots provides the coordinates of the image center. The average separation between the sine spot pairs also provides a plate scale measurement: 6.5 pixels/resel, sufficiently oversampled for the SCC fringes.
\subsection{Interaction and Command Matrices}
\label{sec: im}
Our wavefront calibration and reconstruction procedure largely follows the standard methods developed for calibrating the SCC, e.g., as in Ref. \citenum{fast_spie18}. We briefly summarize these steps here, specifically noting additional setup-specific subtleties we have implemented.

To generate the interaction matrix (IM), sine and cosine waves are applied to the MEMS for every spatial frequency greater than 5 c/p within the DM control region at intervals of 1 c/p. Lower order Fourier modes (i.e., those that are close to or within the FAST FPM inner working angle) would increase non-linearities by modulating the pinhole PSF throughput for only those modes. To minimize aliasing effects we limit the highest MEMS spatial frequency along the x direction to 12 c/p, below the theoretical 14.5 c/p maximum\cite{fast_spie18,fast_spie20}. As described in \S\ref{sec: fast_sum}, every SCC image for the corresponding Fourier mode is operated on by a Fourier transform, a pre-computed binary mask to isolate the fringes (realized as sidelobes in the image MTF), and an inverse Fourier transform to return to the (now complex-valued) image plane, called the $I_-$ plane. For each DM Fourier mode, we compute the $I_-$ pixels values (real and imaginary) within a user-defined pre-computed ``dark hole'' (DH) region (i.e., which sets the 12 c/p maximum spatial frequency limit along the x direction described above) for a differential image (i.e., SCC image with Fourier mode applied minus SCC image with best flat applied). As in Ref. \citenum{fast_spie20} we also implement an additional binary mask before storing the computed $I_-$ for a given mode in the interaction matrix, which isolates the given image-plane sine spot within a 2 $\lambda/D$ radius of it's central location and sets all other pixel values within the DM control region equal to zero, minimizing cross-talk between modes. Looping through all Fourier modes, these $I_-$ pixel values are then stored in a 456$\times$27936 ``reference vector'' which we will call $A$ (456 is the number of Fourier modes; 27936 is the number of different pixel values---both real and imaginary---within the half DH region of $I_-$). The 456$\times$27936 command matrix (CM), $C$, generates a a 456$\times$1 vector of least-squares coefficients, $c$, for each DM Fourier mode given an input 27936$\times$1 (i.e., vectorized) $I_-$ target image, $t$, via $c=C\cdot t$. The CM is computed by $(A\cdot A^T)^{\dagger} \cdot A$, where $\cdot$, $^T$, and $^\dagger$ represent matrix dot product, transpose, and pseudo inverse (see next paragraph) operators, respectively. Finally, a 1024$\times$456 vector of the Fourier modes applied in DM command units, $F$, is used to convert to DM command space, $D$, via $D=F\cdot(-c)$, producing a 1024$\times$1 matrix that can be reshaped to the 32$\times$32 DM actuator dimensions.\footnote{In a future optimized real-time code $F$ will be incorporated into the CM to require only one real-time matrix multiplication instead of two.}

We compute the pseudo inverse via singular value decomposition (SVD), choosing the lowest SVD cutoff value (i.e., the lowest number of principle components propagated through to the CM) which sufficiently reconstructs individual Fourier modes applied on the DM without cross-talk from other modes. Note that although with a more conventional zonal-based pupil plane WFS where the SVD cutoff is used to minimize the propagation of waffle modes onto the DM, this waffle mode propagation is less relevant for a focal plane WFS, since a DH is inherently spatially filtering modes that the DM controls, of which the highest spatial frequency waffle mode is not include in the above-mentioned algorithmic filter applied to $I_-$. However, we found that selecting too many Eigen modes (i.e., too low of an SVD cutoff value) can still cause greater loop stability issues, potentially in enabling the included Fourier components of such a DM waffle mode to still be projected through to the CM, which is why we used the minimal number of Eigen modes to enable sufficient reconstruction of all desired Fourier modes.
\section{CLOSED-LOOP OPERATIONS}
\label{sec: closed_loop}
This section is split into two, first presenting results and analysis of correction only quasi-static errors from optical surfaces in the system and in-air bench stability (\S\ref{sec: quasi_stat}) and next in closing the loop on simulated AO residual turbulence with the MEMS (\S\ref{sec: dynamic}). The latter is an important milestone for focal plane wavefront sensing technologies, demonstrating the potential of FAST to be implemented as a second stage AO WFS. Common to both sections, our current Python-based real-time code runs at approximately 50 Hz with 1 frame of system latency, set by a user-defined 20 ms pause at each iteration to enable FAST images (at the end of the iteration) to see the effect of DM commands applied (at the beginning of the iteration); we found that pauses smaller than this amount could prevent this sequential dependence from functioning properly, although we do not expect this to remain an issue after a planned future migration to the C-based Keck RTC framework (see \S\ref{sec: setup}).
\subsection{Quasi-static}
\label{sec: quasi_stat}
\begin{figure}[!h]
    \centering
    \begin{subfigure}[b]{0.21\textwidth}
        \includegraphics[width=\textwidth]{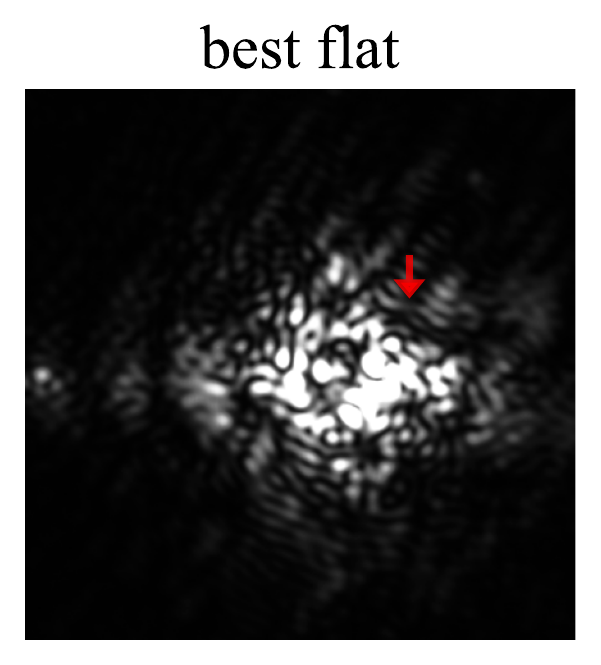}
        \vspace*{0.01pt}
        \vspace*{-6.0pt}
        \caption{}
    \end{subfigure}
    \hfill
    \begin{subfigure}[b]{0.33\textwidth}
        \includegraphics[width=\textwidth]{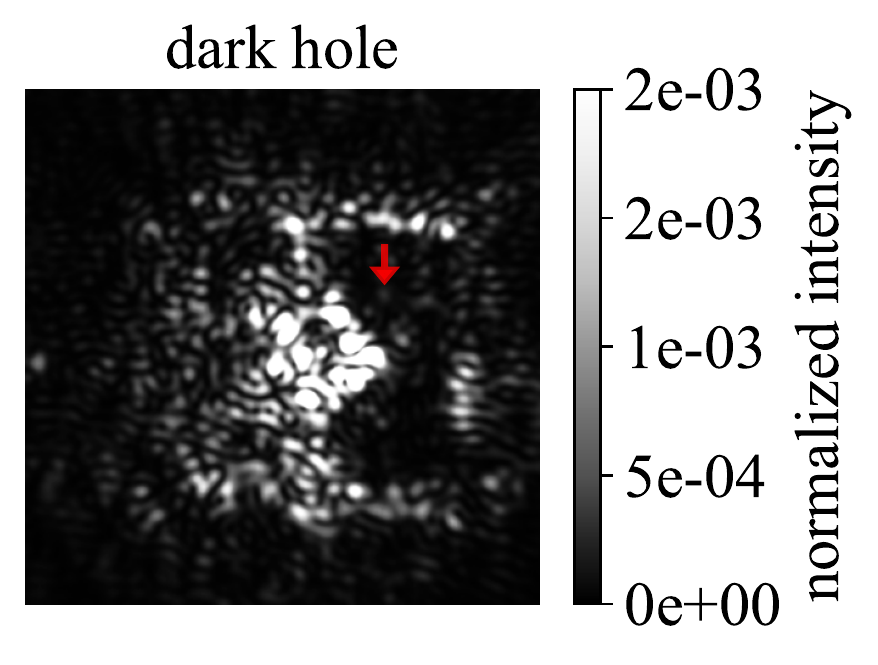}
        \caption{}
    \end{subfigure}
    \hfill
    \begin{subfigure}[b]{0.4\textwidth}
        \includegraphics[width=\textwidth]{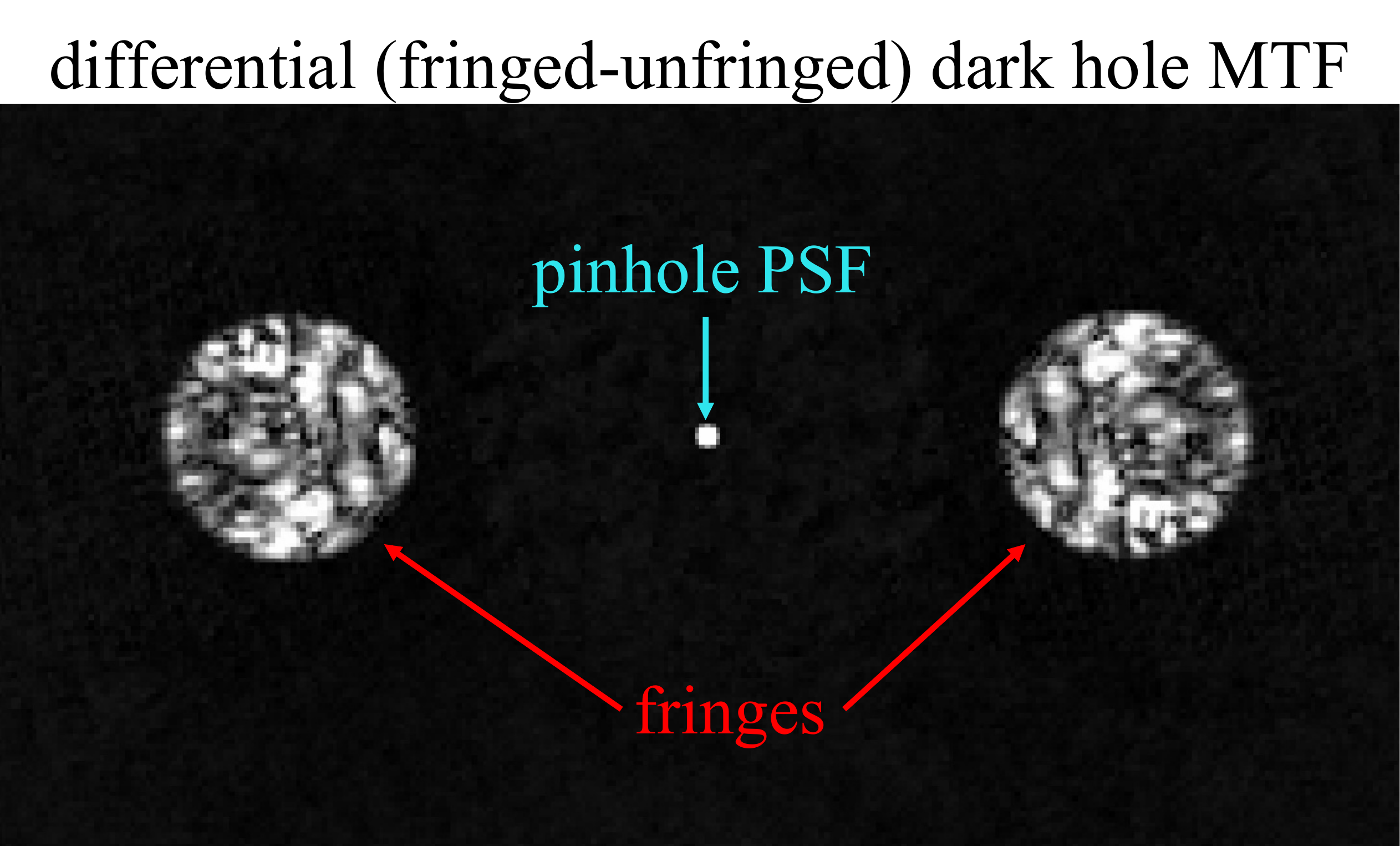}
        \caption{}
    \end{subfigure}
    \hfill
    \begin{subfigure}[b]{0.35\textwidth}
        \includegraphics[width=\textwidth]{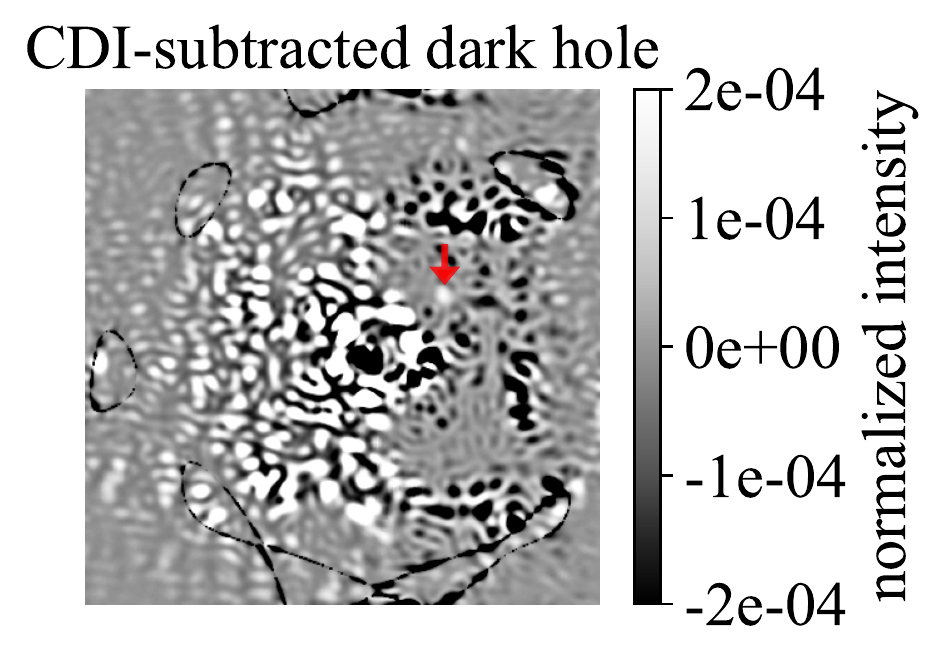}
        \caption{}
    \end{subfigure}
    \hfill
    \begin{subfigure}[b]{0.63\textwidth}
        \includegraphics[width=\textwidth]{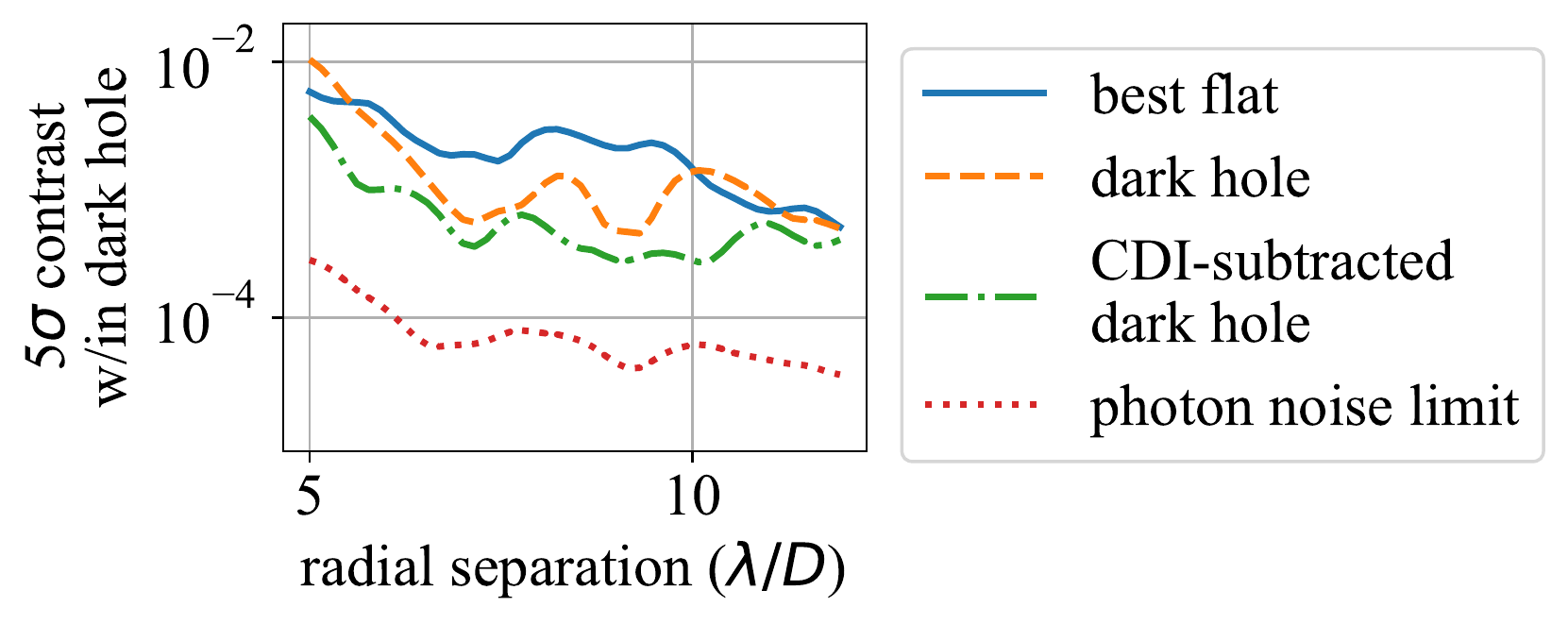}
        \caption{}
    \end{subfigure}
    \caption{Speckle correction/subtraction results of quasi-static aberrations on SEAL, obtained with the Lyot stop optical chopper running/synchronized to the FAST images as described in \S\ref{sec: setup}. (a) FAST image before DM correction. (b) The same image after DM correction, generating a dark hole on the right side of the star. (c) Differential MTF of the fringed and unfringed images used to generate b, illustrating in the Fourier plane the fringe ($2\sqrt{I_S I_R}M$) and pinhole PSF ($I_R$) image components. (d) Resultant output after CDI subtraction of b, revealing the planet (whose position is indicated by a red arrow in panels a, b, and d). (e) Contrast curves for panels a, b, and d, as well as the fundamentally deepest achievable photon noise limit (see the text for more detail). Panels a, b, and d---obtained with the optical chopper running---show un-fringed (i.e., pinhole blocked: $I_S+I_P$) image components on an intensity scale normalized to the off-axis peak stellar flux as described in \S\ref{sec: calib_ini} (a and b are set to the same scale; d is a different scale, as shown, to see the planet source).}
    \label{fig: stat_imas_and_ccurves}
\end{figure}
\begin{figure}[!h]
    \centering
    \includegraphics[width=0.62\textwidth]{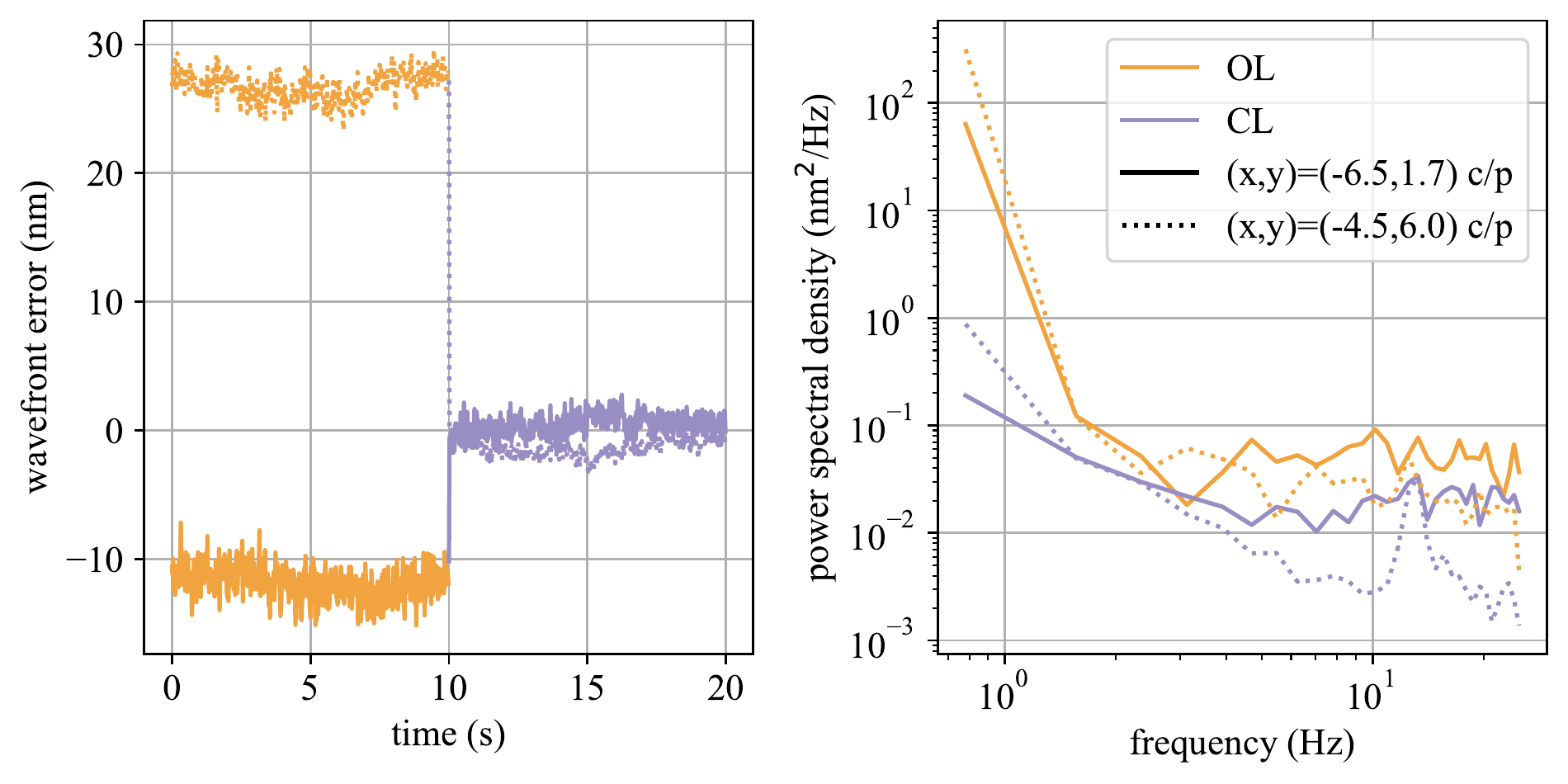}
    \caption{SEAL temporal stability before and after closing the loop on only quasi-static turbulence, showing the time series on the left (with the loop closing---immediately applying the previously determined quasi-static dark hole DM commands used to generate Fig. \ref{fig: stat_imas_and_ccurves}b instead of a ``cold start''---at t=10s) and corresponding temporal PSDs on the right, shown for two different Fourier modes (labeled by different solid and dotted line styles). Open and closed loop (OL and CL, respectively) are distinguished by the orange and purple colors, respectively.}
    \label{fig: static_turb}
\end{figure}
Fig. \ref{fig: stat_imas_and_ccurves} shows images and contrast curves of speckle correction of quasi-static aberrations within the SEAL system, including correction made by DM control and chopper-enabled CDI. 
Contrast curves in panel e, calibrated to the peak off-axis stellar flux as in \S\ref{sec: calib_ini}, compute five times the standard deviation within a given radial annulus that is also within the half DH region on the right hand side of the image. 

CDI processing is enabled by the optical chopper, which as shown in Fig. \ref{fig: stat_imas_and_ccurves}c provides $I_R$ after isolating the central MTF peak and then performing an inverse Fourier transform back to the image plane. Isolating a single side lobe of the MTF in panel b and then performing an inverse Fourier transform provides the complex-valued $I_-$, for which its amplitude is $|I_-|=\sqrt{I_S I_R}$. $I_S$ can then be reconstructed by the two Fourier-filtered terms: $I_S=|I_-|^2/I_R$, which can then be subtracted from the un-fringed image, $I_S+I_P$, to leave $I_P$: $(I_S+I_P)-|I_-|^2/I_{-}=I_P$. Clearly panel d shows that there is more than just $I_P$ left in the image, which is to be expected due to the impact of detector and photon noise, but this fundamental limit (i.e., as defined in Ref. \citenum{fast18}, accounting for the impact of noise propagation through the CDI Fourier filtering algorithms) is about 10x lower, as shown in panel e.\footnote{This photon noise limit is computed by a stack of consecutive differential un-fringed images (i.e., minimizing the impact of wavefront evolution contributing to speckle noise, instead just isolating the effects from detector and photon noise) from the chopper sequence used to generate the image in Fig. \ref{fig: stat_imas_and_ccurves}b, then further processed accounting for noise propagation through our Fourier filtering algorithms as in Ref. \citenum{fast18}.} Initially the reconstructed $I_S=|I_-|^2/I_R$ term is about 2-3x dimmer than the measured $I_S$ term, requiring normalization by standard deviation within the central 5 $\lambda/D$ of the star (i.e., within the FPM inner working angle, preventing any bias from the exoplanet signal). However, even after such normalization, producing the green dashed dotted curve in Fig. \ref{fig: stat_imas_and_ccurves}e, this CDI-subtracted contrast curve is well above the photon noise limit (red dotted) curve. Despite these clear gains enabled by CDI, such discrepancies should be further explored in future work. 

Fig. \ref{fig: static_turb} shows the temporal effects of closing the loop on quasi-static turbulence.
Temporal power spectral density (PSD) plots are generated using both a Hanning window and segment averaging in the time domain to suppress noise before converting to the Fourier domain. With the loop closed, an integrator gain of 0.05 is applied, which is necessary to prevent the loop from going unstable; as shown in Fig. \ref{fig: static_turb}, open loop PSDs are flat all the way down to $\sim$2 Hz, showing a high level of bench stability for which no temporal wavefront evolution signal is detectable above this frequency, meaning an integrator with a 0 dB bandwidth greater than 2 Hz (running at 50 Hz with a 1 frame delay and a pure integrator gain of 0.05, the theoretical 0 dB bandwidth is 0.9 Hz) will only amplify noise/increase loop instabilities. Despite this lack of a detectable wavefront evolution at high temporal frequencies, Fig. \ref{fig: static_turb} still does show that the system is better stabilized at all temporal frequencies (i.e., the ``CL'' purple curve is below the ``OL'' orange curve across all frequencies for both Fourier modes). Integrated temporal WFE between 2 and 25 Hz for open and closed loop are respectively 1.1 and 0.7 nm rms for the solid line Fourier mode and 0.8 and 0.4 nm rms for the dotted line Fourier mode,  demonstrating the ability for FAST to improve quasi-static bench stability in closed vs. open loop.
\subsection{AO Residuals}
\label{sec: dynamic}
\begin{figure}[!h]
    \centering
    \begin{subfigure}[b]{0.23\textwidth}
        \includegraphics[width=\textwidth]{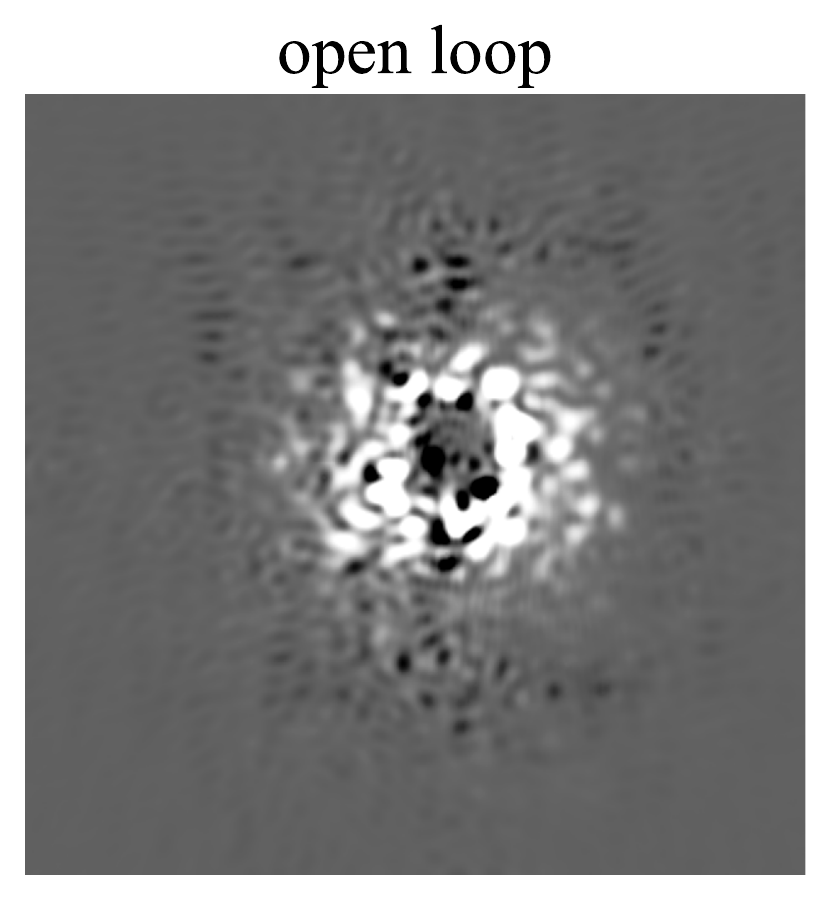}
        \vspace*{0.01pt}
        \vspace*{-10pt}
        \caption{}
    \end{subfigure}
    \hfill
    \begin{subfigure}[b]{0.34\textwidth}
        \includegraphics[width=\textwidth]{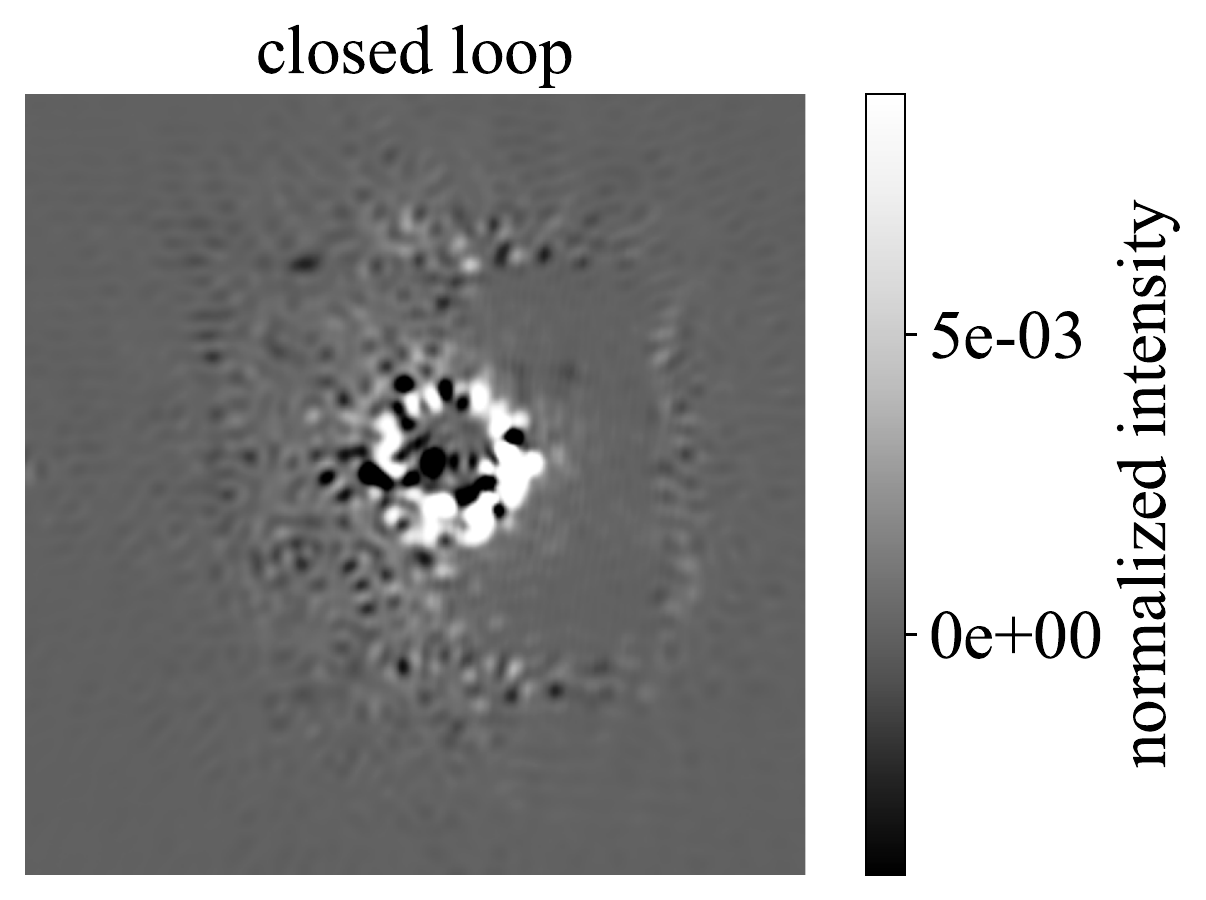}
        \caption{}
    \end{subfigure}
    \hfill
    \begin{subfigure}[b]{0.37\textwidth}
        \includegraphics[width=\textwidth]{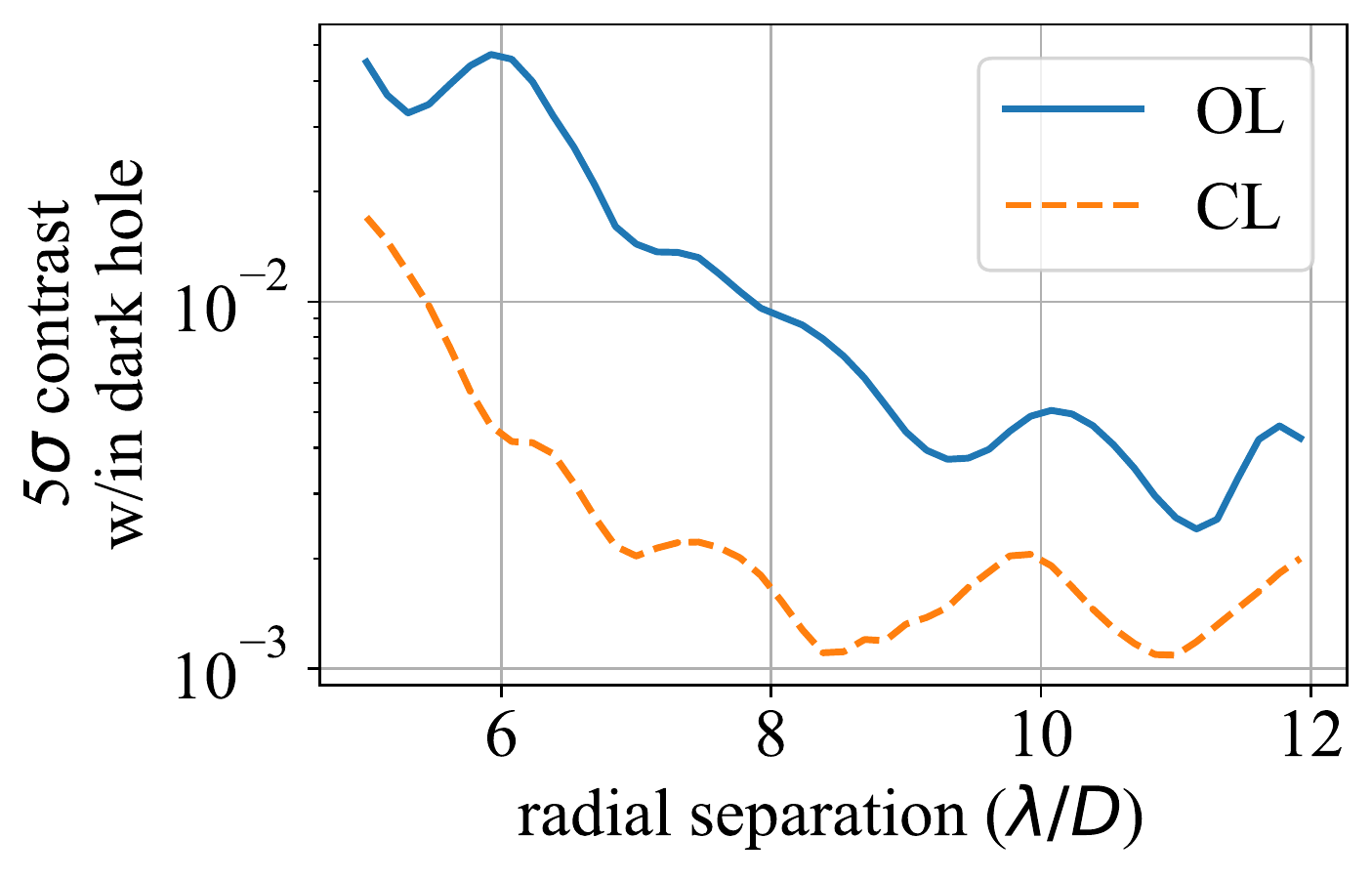}
        \caption{}
    \end{subfigure}
    \caption{Stacked differential un-fringed FAST images (i.e., target image - best flat image from Fig. \ref{fig: stat_imas_and_ccurves}b) for open and closed loop (OL and CL for panels a and b, respectively) correction of evolving AO residuals applied on the DM over a 20s sequence running at 50 Hz% (5s stack for OL, 14s stack for CL)
    . (c) Corresponding contrast curves for a and b.}
    \label{fig: dynamic_imas_and_ccurves}
\end{figure}

In this section we present an important milestone for focal plane wavefront control technologies, demonstrating FAST correction of AO residual speckles in real-time. An evolving AO residual wavefront is applied on the MEMS in open loop and then corrected by FAST in closed-loop with the following parameters assumed/implemented:
\begin{itemize}
    \item Spatial WFE is normalized to 50 nm rms between 3 and 16 c/p assuming $\lambda=1.6\mu$m with a -2 spatial PSD power law. This is equivalent to a 133 nm rms phase screen normalized from 0.5 c/p (i.e., focus; tip/tilt removed) to 16 c/p. However, in our case spatial frequencies less than 3 c/p are set to zero. This removal of low order aberrations is intentional; FAST low order capabilities have been simulated\cite{fast_spie20} and related SEAL testing is currently ongoing, planned for a future separate publication (Sangupta, A, Gerard, B et al., in prep.; also see \S\ref{sec: conclusion}). For this reason we are only focusing on mid-to-high order AO residual correction in this manuscript.
    \item A 10m diameter telescope, 1 m/s ground layer wind speed is assumed (see below for further discussion).
    \item As discussed at the beginning of \S\ref{sec: closed_loop}, our Python real-time code limits operational frame rate to 50 Hz with a 1 frame delay. For this reason we do not use the optical chopper here, which would further limit loop speed to 25 Hz. This limited loop speed is the main reason why a 1 m/s wind speed is required, although a more realistic 1 ms atmospheric lag\cite{fast_spie18} is applied at the time of DM correction. Regardless, future more-optimized real-time code running at kHz speeds will enable (1) closed-loop correction on more typical wind speeds of order 10 m/s and (2) using the optical chopper to provide further CDI subtraction of FAST-corrected atmospheric speckles.
    \item We use a leaky integrator controller, with a leak of 0.9 and gain of 0.4 for all modes. This leak value is particularly lower than normal but we found necessary to ensure closed-loop stability, which we discuss further in \S\ref{sec: stability}. We use the same methods as in Ref. \citenum{fast_spie18} to compute the theoretical rejection transfer function given these parameters.
    \item Closed-loop target SCC images use differential images (with respect to Fig. \ref{fig: stat_imas_and_ccurves}b) dotted with the command matrix to generate DM commands (i.e., instead of absolute speckle correction, preventing achievable closed-loop contrasts deeper than the static DH but enabling faster convergence to closed-loop residual levels).
\end{itemize}
With this setup implemented, Fig. \ref{fig: dynamic_imas_and_ccurves} shows a comparison between long exposures of open and closed-loop FAST correction of evolving AO residuals, clearly illustrating the benefit of closed loop operations at all spatial frequencies.

\begin{figure}[!h]
    \centering
    \includegraphics[width=0.85\textwidth]{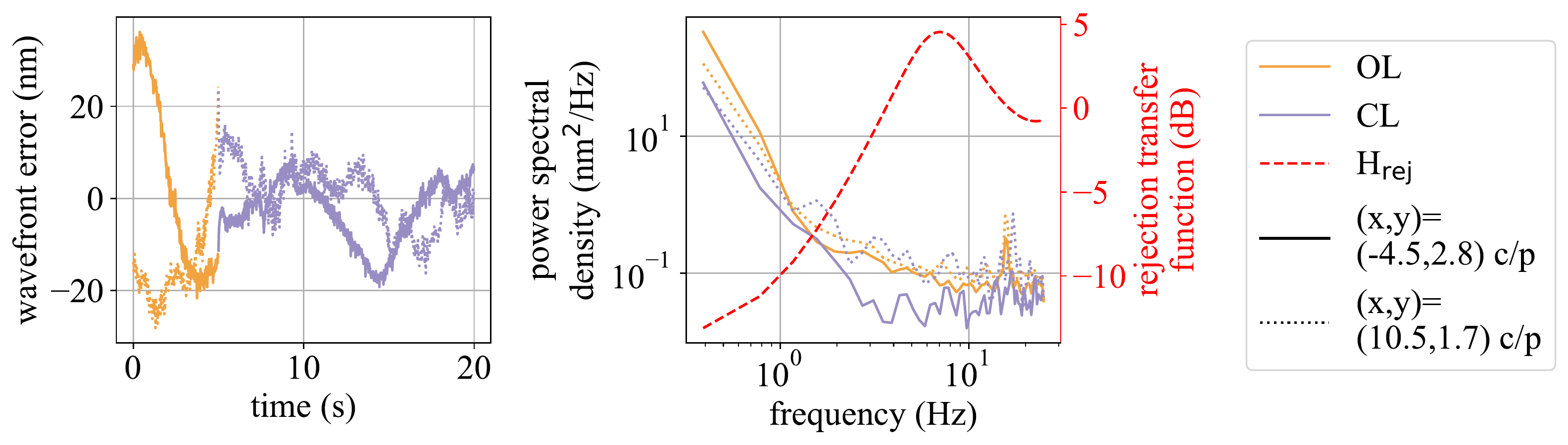}
    \caption{Time series (left panel) and corresponding temporal PSDs (right panel) for evolving AO residuals applied on the MEMS, both for open loop (orange) and closed loop (purple), and shown for one lower and one higher order Fourier mode (solid and dotted lines, respectively). The right y-axis shows the corresponding rejection transfer function (H$_\mathrm{rej}$) for the same displayed frequency range and using the closed-loop and controller parameters presented at the beginning of this section.}
    \label{fig: atm_t}
\end{figure}

Fig. \ref{fig: atm_t} further supports this gain, comparing time series and corresponding temporal PSDs between open and closed loop operations for two Fourier modes.
The lower order Fourier mode (solid line) shows a clear gain of temporal WFE (i.e., purple curve below the orange curve), particularly at frequencies between about 2 - 10 Hz (integrating open and closed loop PSDs between 2 - 10 Hz yields 0.9 and 0.5 nm rms, respectively). Interestingly, the higher order Fourier mode shows a minimal gain in closed-loop at any temporal frequency (integrating over the same region for open and closed loop yields 1.0 and 1.1 nm rms, respectively). Although Fig. \ref{fig: dynamic_imas_and_ccurves}c shows a smaller gain in closed vs. open loop at higher separations in the coronagraphic image (i.e., higher order DM Fourier modes), in principle this factor of $\sim$2 contrast gain at those separations should correspond to a 2 times gain in  temporal PSD space, which could simply require a longer time series so noise averaging can illustrate this relatively small gain that is hard to detect in the 20s data set used to generate Fig. \ref{fig: atm_t}.
\section{DISCUSSION AND FUTURE WORK}
\label{sec: discussion}
\subsection{Loop Stability}
\label{sec: stability}
\begin{figure}[!h]
    \centering
    \includegraphics[width=0.8\textwidth]{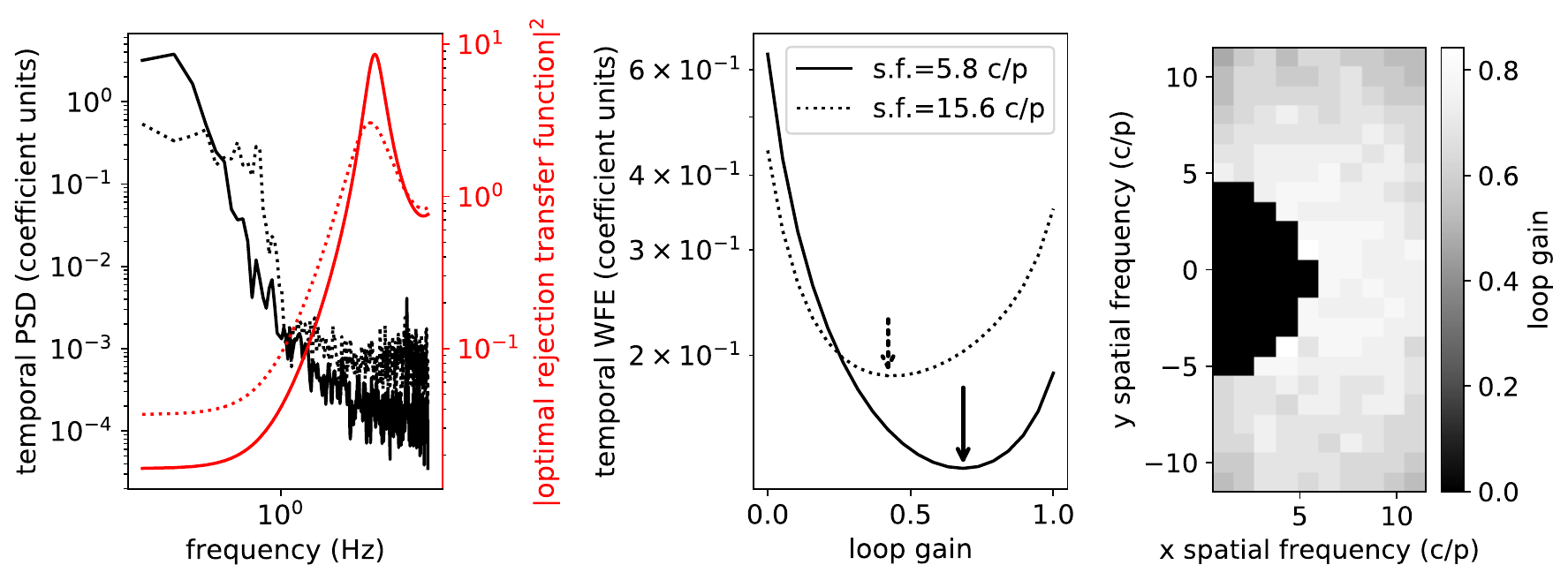}
    \caption{Fourier modal gain optimization analysis. Left panel: open loop temporal PSDs and corresponding square modulus of the optimal rejection transfer function (left and right y-axis, respectively), shown for a lower and higher order Fourier mode (solid and dotted lines, respectively, with ``s.f'' in the middle panel legend---applying also to the left panel---denoting radial spatial frequency from the star). Middle panel: integrated WFE over all temporal frequencies as a function of leaky integrator gain (with the leak set to 0.9 for all Fourier modes). Right panel: optimal leaky integrator gain for all controlled Fourier modes (the gain is set to zero within 5 $\lambda/D$ of the star because these modes are intentionally not controlled, as described in \S\ref{sec: im}).}
    \label{fig: gopt}
\end{figure}

We also tried to implement FAST Fourier modal gain optimization as simulated in Ref. \citenum{fast_spie18}, but ultimately we were not able to generate a stable closed-loop; instead, a constant (lower) gain as presented in \S\ref{sec: dynamic} provided the best results over a time-averaged sequence. Regardless, Fig. \ref{fig: gopt} illustrates our efforts to implement this approach.
As in Ref. \citenum{fast_spie18}, using the AO transfer functions from Ref. \citenum{jp} for a leaky integrator with the system parameters listed in \S\ref{sec: dynamic}, a 45 degree phase margin (i.e., a transfer function metric to ensure AO loop stability) is reached at a gain of 0.6, which is around the minimum gain found for the highest spatial frequencies in Fig. \ref{fig: gopt}; it is therefore not surprising that our modally-optimized gain approach goes unstable compared to a constant lower gain of 0.4. It is also possible that additional non-linear affects not accounted for in the standard AO transfer functions from Ref. \citenum{jp} contribute additionaly to instability (i.e., further lowering the phase margin), such as saturated and/or bad actuators (see below)% (see \S\ref{sec: bad_actuators} below for further discussions)
. Furthermore, compensation of amplitude errors (i.e., scintillation, instrumental, and/or diffraction, although only the latter two are present in our laboratory setup used in this work), which inherently un-flatten the DM, are also additional factors unique to focal plane WFSs that could cause further instabilities.

\begin{figure}[!h]
    \centering
    \begin{subfigure}[b]{0.14\textwidth}
        \includegraphics[trim={0 0 3.8cm 0},clip, width=\textwidth]{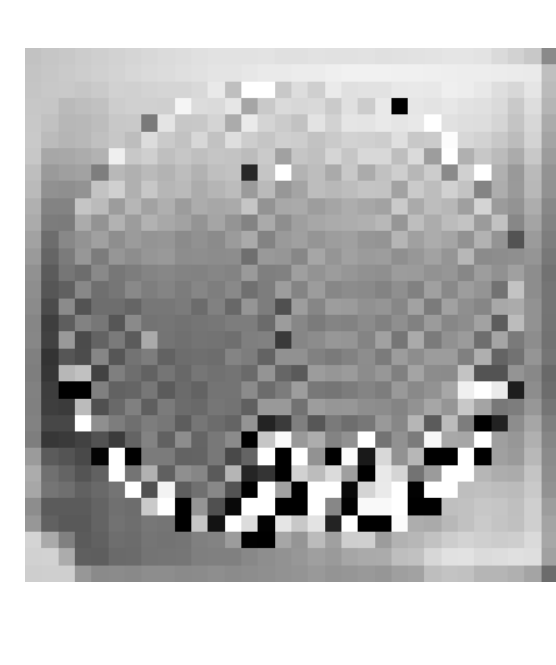}
        \vspace*{-15pt}
        \caption{}
    \end{subfigure}
    \hfill
    \begin{subfigure}[b]{0.23\textwidth}
        \includegraphics[width=\textwidth]{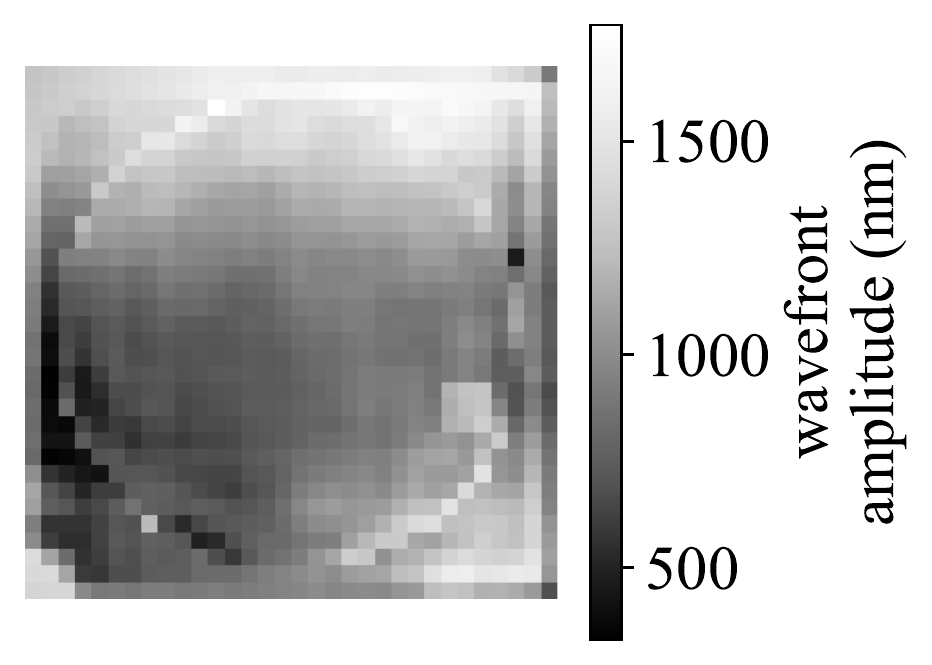}
        \caption{}
    \end{subfigure}
    \hfil
    \begin{subfigure}[b]{0.33\textwidth}
        \includegraphics[width=\textwidth]{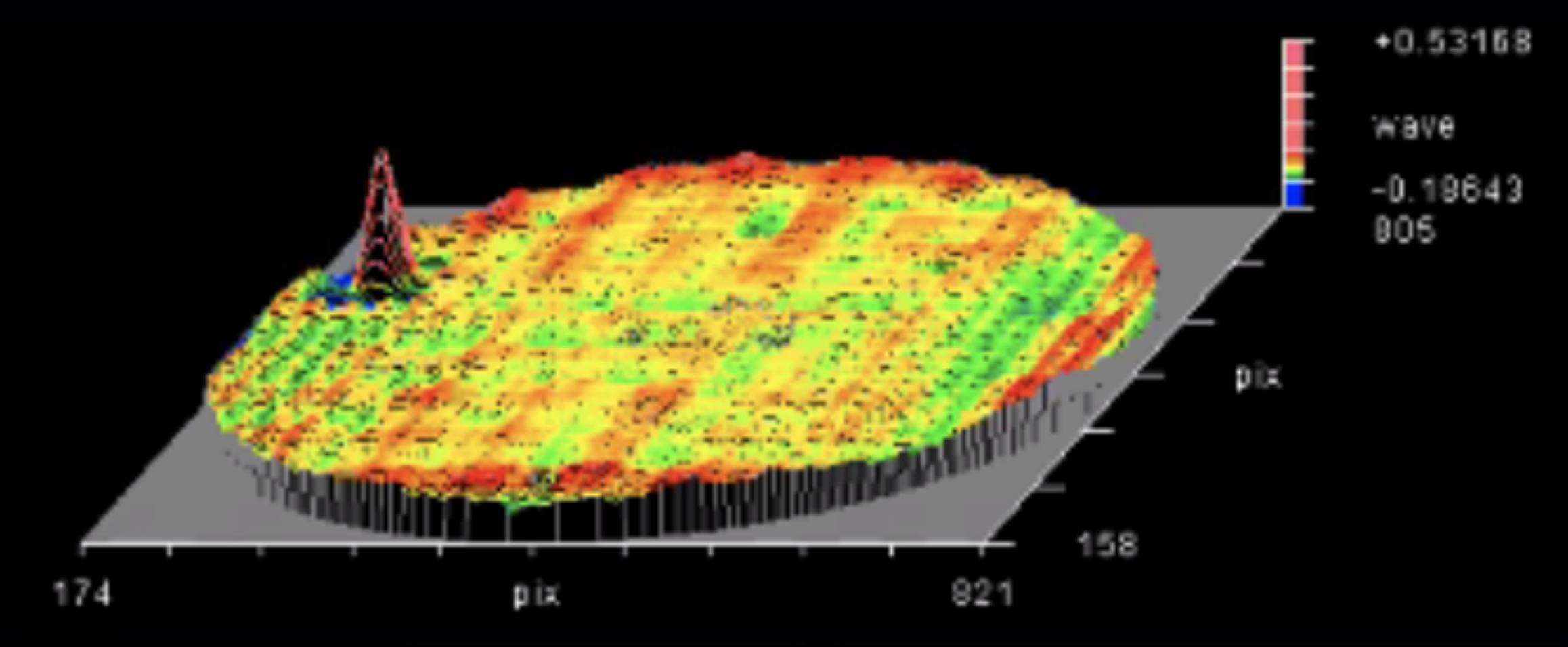}
        \caption{}
    \end{subfigure}
    \hfill
    \begin{subfigure}[b]{0.28\textwidth}
        \includegraphics[width=\textwidth]{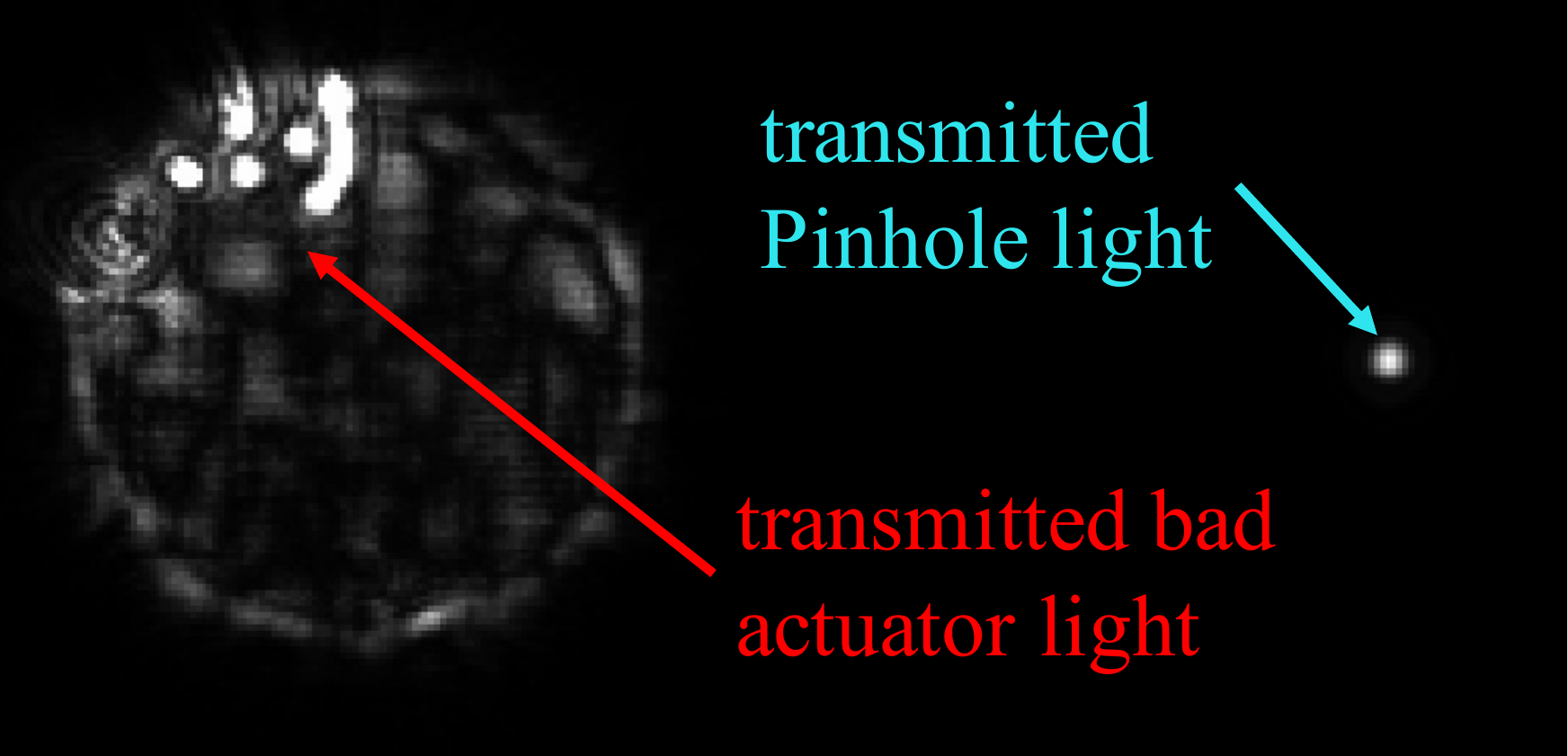}
        \caption{}
    \end{subfigure}
    \caption{(a) MEMS actuator commands, converted to calibrated wavefront units (see \S\ref{sec: calib_ini}), applied to generate the dark hole in Fig. \ref{fig: stat_imas_and_ccurves}b and d. (b) MEMS commands applied to flatten the wavefront, generating the image in Fig. \ref{fig: stat_imas_and_ccurves}a. (c) Zygo Interferometer image from April 2021 using MEMS commands from panel b, showing a single bad actuator at the edge of the system aperture (i.e., outside the 29 actuator system diameter). (d) Coronagraphic pupil image from August 2021 produced from the same MEMS commands as in panels b and c, illustrating the presence of additional pinned actuators that are not blocked by a Lyot stop. Note that DM commands in panel b are flipped left-right and up-down with respect to panels c and d.}
    \label{fig: dmc_dh}
\end{figure}

Saturated actuators are particularly concerning in terms of factors that could be causing bench instabilities, which are known for MEMS to be highly non-linear at these stoke limits\cite{poyneer_mems}. Fig. \ref{fig: dmc_dh} shows the MEMS actuator commands applied to generate the best flat and static DH in Fig. \ref{fig: stat_imas_and_ccurves}a and b, respectively, and also illustrates the best flat degradation over the course of the testing and development included in this paper (April - August 2021).
These saturated and/or pinned actuators in Fig. \ref{fig: dmc_dh} both (1) do not apply simulated turbulence because no further stroke can be applied and (2) may be contributing to overall closed-loop system instabilities (including nearby actuators that are close to being saturated/pinned due to the aforementioned non-linear properties at these stroke limits).
\subsection{Tilted Pinhole PSF}
\label{sec: tilted_pinhole_PSF}
For many results presented in this paper we noticed an unusual behavior in our SCC images with the MEMS aligned in the system: the pinhole PSF center appears off-set from the unfringed coronagraphic image center. In this section we discuss a rationale as to why this might be happening. 

Note that during the ``by-eye'' alignment process described in \S\ref{sec: fpm_alignment}, we noticed an apparent degeneracy between FPM position angle (PA) and x-y positioning alignment, with both moving the off-axis spot to a different location on the Lyot stop; although we want the optimal alignment of the star centered on the FPM, small x-y positioning adjustments off-of alignment move the location of the Lyot plane off-axis pupil's peak flux in x and y, therefore allowing compensation of slight FPM PA alignment offsets with a x-y FPM position adjustment (i.e., where by eye the peak intensity of the FAST off-axis pupil is still centered on the Lyot stop pinhole). This degeneracy was easier to resolve for the characterization done in \S\ref{sec: uoa_fpm} because a fold mirror replaced the MEMS, enabling an overall much lower system WFE at the FPM plane and illustrating a degraded contrast when such a sub-optimal compensation approach was applied. The FPM characterization work in \S\ref{sec: uoa_fpm} therefore shows both an aligned pupil and focal plane image, where in the former the off-axis pupil is centered on the Lyot stop pinhole and in the latter both the contrast is minimized and the pinhole PSF and coronagraphic image are centered with respect one another. However, with the MEMS in the beam (i.e., in \S\ref{sec: closed_loop}) it was much harder to resolve this degeneracy, which may account for the fact that although the pupil images seem aligned, the focal plane images still show a tilt of the pinhole PSF with respect to the coronagraphic image. In other words: sub-optimal FPM PA alignment with respect to the Lyot stop pinhole PA that is then compensated for by sub-optimal x-y FPM positioning alignment causes a tip/tilt phase shift across the Lyot stop pinhole which is not present across the main coronagraphic pupil, resulting in a tilted pinhole PSF with respect to the unfringed coronagraph image center. 

In order for future instruments to resolve this ambiguity, ideally the FPM would have an automated stage for PA adjustment, where in addition to DM tip/tilt, FPM PA can also be iterated over in a procedure to search for optimal alignment of all three degrees of freedom, thereby ensuring that no differential pinhole PSF offset is present. We are investigating options to include this in both our laboratory setup and future FAST instruments.  
\subsection{Future In-House FAST Coronagraph Mask Fabrication and Testing}
\label{sec: future_ucsc_fpms}
We are considering several future avenues of exploration for in-house coronagraph mask developments, including improvements on reflective design fabrication capabilities and new coronagraph designs to fabricate and test.

The reflective 4-level Ti masks are not ideal as Ti is not optimally reflective at visible or infrared wavelengths, which is not an issue for our lab testing with a bright light source but not feasible to implement in an on-sky system. Although our current facilities prevent us from depositing layered aluminum (the development step, which washes away exposed photo-resist, would etch aluminum deposited by a previous layer), we have already tested a final Al coating step after all layers of the Ti mask are fabricated; however, subsequent characterization (via SEAL lab tests as well as optical profilometry and scanning electron microscopy) showed that this Al coating added considerable figure error to the surface profile (on the order of a few hundred nm rms). Accordingly, we are in the process of developing modifications to the Al coating step to minimize this added figure error. We are also considering testing designs with more than four layers, since a four-level design is not ideal in terms of achievable contrast and exoplanet throughput but Ref. \citenum{scalar_vortex} has showed that a six-level design enables significant improvement, with the downside of adding additional alignment steps between layers increasing risk of greater figure errors in the final fabricated mask.

We will also soon be testing the FAST-compatible, multi-reference SCC (MRSCC), which adds multiple pinholes in the Lyot stop to be more robust to fringe smearing effects over a larger bandpass\cite{mrscc}. Ref. \citenum{mrscc} has already demonstrated laboratory broadband dark hole generation with the classical SCC, and in Ref. \citenum{fast_phd} we initially proposed the concept of combining this with FAST, adding multiple tilt angles to a TG or TGV mask to redirect the PSF core at the FPM into three different pinholes, enabling both high-speed and broadband wavefront control with FAST. A forthcoming paper will present SEAL testing results of this concept, using the same in-house 4-level Ti metal deposition fabrication capabilities as described above.
\section{CONCLUSION}
\label{sec: conclusion}
Exoplanet imaging speckle subtraction via focal plane wavefront control and/or CDI is a promising technology that has the potential to enable current and future instruments to detect and characterize lower mass, closer in, and/or older exoplanetary systems than is currently possible. However, one issue thus far in making such technology operational on-sky has been achievable temporal bandwidth, due to continual speckle evolution from atmospheric and quasi-static effects. We have developed a technique to mitigate this bandwidth issue, called the Fast Atmospehric Self-coherent camera Technique (FAST), for which in this manuscript we present laborabory validation on the Santa cruz Extreme AO Laboratory (SEAL) testbed. Our main findings are as follows:
\begin{enumerate}
    \item We have designed and tested two fabricated FAST coronagraph masks---one Tip/tilt Gaussian mask from U. of Alberta's nanoFAB laboratory and one Tip/tilt Gaussian Vortex mask from UCSC's W.M. Keck Center for Nanoscale Optofluidics--- both demonstrating sufficient contrast and fringe visibities for FAST ground-based operations.
    \item We tested FAST quasi-static DH generation---including with CDI-based post-processing---and closed-loop temporal stability, demonstrating 5$\sigma$ contrasts within the DH of around 5e-4 (Fig. \ref{fig: stat_imas_and_ccurves}e) and temporal stabilities per Fourier mode below 1 nm rms between 2 and 25 Hz (Fig. \ref{fig: static_turb}).
    \item We tested FAST correction of AO residual speckles, demonstrating a clear gain of such a high speed ``second stage AO'' capability using a focal plane wavefront sensor, showing a factor of 2 - 10 times contrast gain, depending on radial separation (Fig. \ref{fig: dynamic_imas_and_ccurves}).
\end{enumerate}
These testing results clearly illustrate the benefit of running fast focal plane wavefront control and CDI technology at high speed. However, the developments presented here are by no means comprehensive, instead opening many new avenues of high speed focal plane wavefront control approaches to explore and optimize. In future manuscripts we plan to investigate FAST SEAL testing of many such additional topics (some of which are discussed in \S\ref{sec: discussion}), including low order control and optimization, increasing spectral bandwidth capabilities, optimizing mid-to-high order temporal bandwidth (including predictive control), and real-time integration with a first stage AO correction.
\section*{Acknowledgments}
We gratefully acknowledge research support of the University of California Observatories for funding this research. This work was supported by the W.M. Keck Center for Nanoscale Optofluidics and the Astrophotonics Initiative at UC Santa Cruz. We thank Maaike van Kooten, Renate Kupke, Jules Fowler, and Phil Hinz for comments, suggestions, and discussions that have contributed to this manuscript.

% References
\bibliography{report} % bibliography data in report.bib
\bibliographystyle{spiebib} % makes bibtex use spiebib.bst

\end{document}